\documentclass[twocolumn]{emulateapj}

\usepackage{graphicx}
\usepackage{latexsym,amssymb}
\usepackage{amsmath}
\usepackage{natbib}

\newcommand{\kms}{km\,s$^{-1}$}
\newcommand{\dgr}{$^\circ$}

\newcommand{\pyrGpcc}{yr\,$^{-1}$\,Gpc$^{-3}$}
\newcommand{\Rsun}{R$_\odot$}
\newcommand{\Msun}{M$_\odot$}

\newcommand{\Msunpcc}{M$_\odot$\,pc$^{-3}$}

\newcommand{\Mpcc}{Mpc$^{-3}$}
\newcommand{\nc}{n_\mathrm{c}}
\newcommand{\rc}{r_\mathrm{c}}
\newcommand{\sigc}{\sigma_\mathrm{c}}
\newcommand{\rhoc}{\rho_\mathrm{c}}
\newcommand{\nuenc}{\nu_\mathrm{enc}} 
\newcommand{\nucol}{\nu_\mathrm{col}} 
\newcommand{\nutid}{\nu_\mathrm{tid}} 
\newcommand{\Rmin}{R_\mathrm{min}} 
\newcommand{\Renc}{R_\mathrm{enc}} 
\newcommand{\Rcol}{R_\mathrm{col}} 
\newcommand{\Rtid}{R_\mathrm{tid}} 
\newcommand{\Rdyn}{R_\mathrm{dyn}} 
\newcommand{\xicol}{\xi_\mathrm{col}} 
\newcommand{\rh}{r_h} 
\newcommand{\rt}{r_t} 
\newcommand{\trh}{t_\mathrm{rh}} 
\newcommand{\trhini}{\trh(0)} 
\newcommand{\xides}{\xi_\mathrm{dis}} 
\newcommand{\cini}{c_\mathrm{ini}} 
\newcommand{\tage}{t_\mathrm{age}} 
\newcommand{\Lg}{L_\mathrm{gal}} 
\newcommand{\Phig}{\Phi_\mathrm{gal}} 
\newcommand{\Ls}{L\star} 
\newcommand{\Phigs}{\Phi_\mathrm{gal}\star} 

\slugcomment{Draft \today}

\shorttitle{SGRBs from tidal capture of compact objects in GCs}

\shortauthors{Lee, Ramirez-Ruiz \& van de Ven}

\begin{document}
\title{Short gamma-ray bursts from dynamically-assembled  compact binaries in globular clusters: pathways, rates, hydrodynamics and cosmological setting}

\author{William H. Lee\altaffilmark{1}}

\author{Enrico Ramirez-Ruiz\altaffilmark{2}}

\author{Glenn van de Ven\altaffilmark{3,4}} 

\affil{$^1$Instituto de Astronom\'{\i}a, UNAM, Apdo. Postal 70--264,
  Cd. Universitaria, M\'{e}xico DF 04510; wlee@astroscu.unam.mx}

\affil{$^2$Department of Astronomy and Astrophysics, University
  of California, Santa Cruz, CA 95064; enrico@ucolick.org}

\affil{$^3$Max Planck Institute for Astronomy, K{\"o}nigstuhl 17, 69117
Heidelberg, Germany; glenn@mpia.de}

\affil{$^4$Institute for Advanced Study, Einstein Drive, Princeton, NJ
  08540, USA; Hubble Fellow}

\begin{abstract}
We present a detailed assessment of the various dynamical  pathways leading to the coalescence of compact objects in Globular Clusters (GCs) and Short Gamma-Ray Burst (SGRB) production. We consider primordial binaries, dynamically formed binaries (through tidal two-body and  three-body exchange interactions) and  direct impacts of compact objects (WD/NS/BH). Here we show that if the primordial binary fraction is small, close encounters dominate the production rate of coalescing  compact systems. We find that the two dominant  channels are the interaction of field NSs with dynamically formed binaries, and  two-body encounters. Under such conditions, we estimate the redshift distribution and host galaxy demographics of SGRB progenitors, and find that GCs can provide a significant contribution to the overall observed rate.

Regarding the newly identified channel of close stellar encounters involving WD/NS/BH, we have carried out precise modeling of the hydrodynamical evolution, giving us a detailed description of the resulting merged system. Our calculations show that there
  is in principle no problem in accounting for the global energy
  budget of a typical SGRB. The particulars of each encounter,
  however, are variable in several aspects, and can lead to
  interesting diversity.  First and most importantly, the
  characteristics of the encounter are highly dependent on the impact
  parameter. This is in contrast to the merger scenario, where the
  masses of the compact objects dictate a typical length and
  luminosity scale for SGRB activity. Second, the nature of the
  compact star itself can produce very different outcomes.  Finally,
  the presence of tidal tails in which material will fall back onto
  the central object at a later time is a robust feature of the
  present set of calculations. The mass involved in these structures
  is considerably larger than for binary mergers. It is thus possible
  to account generically in this scenario for a prompt
  episode of energy release, as well as for activity many dynamical
  time scales later.
\end{abstract}


\section{Introduction}
\label{sec:intro}
Until fairly recently, SGRBs were known predominantly as bursts of
$\gamma$-rays, and largely devoid of observable traces at lower
energies. The launch and successful operation of the {\it Swift}
satellite has now enabled the detection and localization of X-ray
afterglows from several events, enabling in turn the study of their
properties at optical and radio wavelengths and the identification of
the host galaxies at cosmological distances
\citep{berger05,fox05,barthelmy05,hjorth05,gehrels05,bloom06,prochaska06}.
The occurrence of a fraction of events among  old stellar populations e.g., of an elliptical
galaxy for GRB 050724, rules out a source uniquely associated with
recent star formation. In addition, no bright supernova is observed to
accompany SGRBs, contrary to what is seen in most nearby long-duration
GRBs. It is clear by now that short and long events are not drawn from
the same parent stellar population, and that SGRBs are far from
standard \citep{nakarrev,gehrels09}. This hints at the underlying possibility
that the progenitor itself may be quite different from burst to burst,
and not entirely restricted to the most widely favoured scenario
involving the merger of close binaries
\citep[e.g.][]{paczynski86,eichler89} containing neutron stars (NS)
and/or black holes (BH).

Since the first evidence from the {\it Uhuru} and
OSO-7 satellites revealed a population of highly luminous low-mass
X-ray binaries (LMXBs) in globular clusters (GCs), it has been noted
that the formation rate per unit mass of these objects is orders of
magnitude higher in GCs than in the Galactic disk
\citep{katz75,clark75}. This discovery stimulated a flurry of
theoretical work into the formation of GCs LMXBs by the processes of
two- and three-body encounters\citep{fabian75,heggie75,hills75}. These
dynamical formation scenarios are a natural explanation for the high
occurrence of LMXBs in GCs since the stellar densities, and hence
encounter rates, are much higher in the cores of GCs than other
regions of the Galaxy \citep{pooley03,heinke06}. In such environments, in fact,
it is unavoidable that many stars undergo close encounters and even
physical collisions, with high probability, within their lifetimes \citep[see e.g.,][]{rosswog09,raskin09}.
It is the interplay between compact stars in such dense environments
and their ability to trigger SGRBs that forms the main topic of this
paper.

In general, forming a compact binary system
requires some mechanism to dispose of enough energy to effectively
bind them. Two such processes have been considered for conditions in
globular clusters \citep{hut83}. The first is the presence of an
additional star which can carry away some energy in kinetic form. This has been
considered by \citet{grindlay06} as a plausible SGRB channel by computing the outcomes of exchange interactions between binaries containing one compact object and a single neutron star in a collapsed GC core. The
second is the loss of energy to internal stellar oscillations, excited
by the tidal forces of one star on the other.  

Here we suggest an alternative and perhaps less restrictive
mechanism for SGRB production related to the second mechanism given above, namely, the tidal capture and collision of compact objects in dense stellar environments. In this new scenario, the compact objects are contained within a globular cluster, and interact directly through close encounters rather than being driven
together by pure gravitational wave emission in existing close binaries.  Event
rates for such interactions within galaxies have been found to be much too low to be of
interest \citep{janka96b} when compared with the GRB event rate, but
they may be frequent enough to have an important effect on their
production in GCs\footnote{\citet{hansen98} have proposed collisions of compact objects with main sequence stars as possibly relevant for the production of GRBs in GCs, although it is not clear how the large baryon loading problem could be circumvented in this case}. In this paper we make a careful assessment of the various dynamical evolutionary pathways involving compact objects in such environments, focusing particularly on the frequency and physical character of the tidal capture of two compact objects that are ultimately capable of leading to a merger and thus powering a gamma-ray burst. We find that they can provide a substantial contribution to the total rate, and compute the evolution of the associated rates with redshift.

Some pressing questions include: When would two passing relativistic
stars capture each other in a bound close orbit? When two neutron
stars collide, does the rapidly-spinning merged system have too much
mass (for most presumed equations of state) to form a single stable
object? If so, the expected outcome after a few milliseconds would
therefore be a spinning BH, orbited by a torus of neutron-density
matter. When a NS collides with a BH, does enough mass remain in the
orbiting debris to catalyze the extraction of energy from the hole at
a rate adequate to power a short-lived GRB?  How do the long tidal
tails thrown out through the outer Lagrange point affect the accretion
stream around the primary star?  Even if the evolution time scale for
the bulk of the debris torus were no more than a second, is enough
mass and energy still available to power the late time flares?  What
is the relative frequency and observable signatures of all these
collision events and how do they compare to {\it Swift} observations?

The structure of this paper is as follows: detailed hydrodynamic
simulations of encounters of compact objects of various types and with
varying impact parameters are presented in Section~\ref{sec:dynamics}
together with a detailed description of the numerical methods and the
initial models; the resulting gravitational wave signals are shown in
Section~\ref{sec:gwaves}; Section~\ref{sec:rates} offers an estimate
of the encounter rate as function of cosmic time, and compares it with
the merger rate of compact binaries. We discuss our findings in
Section~\ref{sec:disc}, and we summarize and conclude in
Section~\ref{sec:ccl}.

\section{Dynamics of tidal capture and disruption}\label{sec:dynamics}

\subsection{Numerical implementation}\label{sec:numerics}

The hydrodynamical calculations shown subsequently have been performed
with the three dimensional Smooth Particle Hydrodynamics (SPH) code
previously used to study merging double neutron star and black
hole--neutron star binaries \citep{monaghan92,lee99a}. The problem
does not allow for simplifications due to symmetry and SPH, being a
Lagrangian scheme lends itself particularly well to this kind of
situation. The tidal tails seen in binary mergers [see
  e.g. \cite{rasio94,lee01,rosswog03}] are a natural outcome in the
present scenario as well, and following their formation and evolution
is one of the main objectives here, which is not possible with
grid--based codes. Self--gravitating spherical stars of a given mass,
$M$, and radius, $R$, in hydrostatic equilibrium with a polytropic
pressure--density relation $P=K\rho^{\Gamma}$, where $K$ and $\Gamma$
are constants, are constructed with $N\simeq 10^5$ fluid elements (SPH
``particles'') and used for the dynamical evolution.  We have
performed test simulations with varying spatial resolution to test for
convergence, using from $N\simeq 10^4$ to $N\simeq 2 \times 10^5$ SPH
particles. We are here mainly interested in the dynamics of the
disruption process for a limited number of dynamical times, and have
found that $N\simeq 10^5$ particles are sufficient for this
purpose. We thus report all simulations at this initial uniform
resolution level. Our calculations are Newtonian, so we cannot model a
true black hole. We merely approximate one as a point mass $M_{\rm
  BH}$, any matter approaching within a Schwarzschild radius $r_{\rm
  Sch}=2GM_{\rm BH}/c^{2}$ being accreted (and its mass added to that
of the hole). Clearly this is a problem that ideally should be treated
in full General Relativity, but a first understanding can be
gained with the use of this approach. We note that previous work on
coalescing black hole-neutron star binaries \citep{lrrg05,rosswog05}
has made use of the pseudo-Newtonian potential of \citet{pw80} for the
black hole, which reproduces the existence and position of a last
stable orbit for test particles in circular orbits around a
Schwarszchild black hole. Doing this thus requires considering fairly
small mass ratios, a condition which is not met in the present set of
calculations, with the additional complication of highly eccentric
orbits for the parabolic encounters of interest. We have hence elected
to keep to a purely Newtonian formulation which, albeit simplistic, is
quantifiably so.

For neutron stars, most equations of state reveal that the radius
varies little over a range of masses. In the case of polytropes, the
mass--radius relation is $R \propto M^{(\Gamma-2)/(3\Gamma-4)}$, so if
the adiabatic index is $\Gamma=2$, the radius is only a function of
the structure constant $K$ and the central density, $\rho_{\rm
  c}$. Our standard ``neutron star'' is thus a spherical,
non--spinning polytrope with $\Gamma=2$, $M_{\rm NS}=1.4 M_{\odot}$
and $R_{\rm NS}=13.4$~km. In order to investigate the effects of a
different compressibility, we have also considered a neutron star with
index $\Gamma=5/3$ of the same mass and radius for certain orbital
parameters, detailed below. We have also considered the case of a
low--mass white dwarf ($M_{\rm WD}=0.5M_{\odot}$) interacting with a
black hole. In this case it is appropriate to use the equation of
state for a cold non--relativistic Fermi gas, $P=K_{\rm n.r.}
\rho^{5/3}$, giving a radius $R_{\rm WD}=1.1 \times 10^4$~km.

Three-dimensional calculations of binary interactions are typically
evolved for only a few tens of milliseconds, and during this short
time large scale gravitational dynamics determine the final state of
the system. Once the initial conditions are set, we use a simple ideal
gas equation of state, where the pressure is given by $P=\rho u
(\Gamma-1)$ and $u$ is the specific internal energy, to follow the
thermodynamics of the gas, with no heating or cooling mechanisms
present. Shocks are allowed to form in the usual numerical way
(through an artificial viscosity), and are the only source of local
dissipation.

The single additional ingredient that is necessary in terms of global
dissipation is the emission of gravitational waves, since it can (and
does) affect the orbital evolution. For binary mergers, the
point--mass approximation in the weak field limit is often used to
compute an effective drag, removing angular momentum and energy from
the system. It is switched off once the stars come into contact, or if
one of them is tidally disrupted. However, for the present set of
calculations on various orbits it is not a good approximation. Rather,
we require an expression for the energy loss rate of extended bodies,
which can be used for more general trajectories (for example initially
parabolic). We thus compute the rate of energy loss as
\begin{equation}
\frac{dE_{\rm GW}}{dt}=\frac{1}{5} \frac{G}{c^{5}} \left[
\frac{d \ddot{I}_{jk}}{dt} \frac{d \ddot{I}_{jk}}{dt} \right],
\end{equation}
where $I_{jk}$ is the traceless mass quadrupole moment.  The first two
time derivatives can be easily calculated without having to perform
two numerical time derivatives explicitly with the use of the
continuity and Euler equations \citep{rasio92,lee99a}. The third
derivative requires numerical treatment, and numerical noise in its
calculation can be reduced by interpolation over neighboring points in
the evolution. We have tested this implementation by computing the
evolution of binaries with large separations, where the orbital decay
time scale is much longer than the orbital period, and find excellent
agreement with the point-mass formula for circular orbits. It is
important to consider this more accurate treatment for parabolic
encounters, since the energy loss rate at periastron can be
substantially larger (by a factor of 2-3, depending on the impact
parameter) than that estimated from the expression for point masses in
circular orbits, leading to different, and overall, faster
encounters. In addition, the effect on the loss rate because of the
formation of tidal bulges on the stars is automatically taken into
account, since it is the full fluid quadrupole moment that is used in
computing the loss rate.

\subsection{Conditions for stellar encounters}\label{sec:ICs}

The stellar velocity dispersion in GCs is insignificant when compared
to that acquired from the mutual gravitational acceleration of two
compact, stellar mass objects as they approach each other. It is thus
reasonable to assume at first that encounters will be parabolic, i.e.,
with vanishing orbital energy, $E$, at infinity. However the impact
parameter, $b$, may vary, and is related to the total orbital angular
momentum, $L$, in the two--body system.

When the separation between the two masses, $a$, is large compared to
their individual radii, $R_{\rm i}$, we may safely consider that they
behave as point particles and compute the orbit accordingly. It is
only when these quantities become comparable that finite--size effects
need to be considered, and a full hydrodynamic calculation must be
performed.  It has been shown for the case of compact binary {\em
  mergers} that tidal effects can de-stabilize the orbit even in the
absence of relativistic considerations at small enough separations
\citep{lai93a,lai93b,rasio94}. The orbital decay rate thus induced is
comparable to that due to gravitational radiation back reaction, and
can even dominate the evolution for the last few cycles. The equation
of state plays a role in the magnitude and growth of this effect, and
for neutron stars it is almost certainly generically important. We
have thus elected to study collisions by placing the stars (whether
they are black holes, neutron stars or white dwarfs) initially at
separations comparable to those used in merger calculations. This
choice is always a trade-off, since using a larger separation will
always be more accurate, at the cost of added computational effort
(during which the components approach each other with little visible
evolution). The point worth noting here is that as a result of tidal
effects, a bulge will form on each star as they approach, but will
slowly fall behind the line joining the centers. At periastron, a
significant lag angle may have developed, and this can have an effect
on the subsequent evolution of the material, through the action of
gravitational torques by the more massive primary on the lighter,
bar--like secondary.  For the present set of calculations we have
chosen to use spherical, non-perturbed stars for initial conditions. A
solution in terms of compressible tri--axial ellipsoids is possible
for the previous evolution \citep{lai93b}, but given our overall
simplifications we believe this will not affect our results
significantly. At the chosen initial separations, the induced tidal
bulges and lag angles are still quite small\footnote{A detailed
  comparison of the differences between the use of spherical stars as
  opposed to tri--axial ellipsoids was carried out in the case of
  mergers \citep{lee00,lee01}. Overall, results differed by less than
  $\approx 5$\% for the equation of state used here with $\Gamma=2$.}.

A convenient way to parameterize the encounters is by comparing the
strength of the tidal field produced by the primary to the self
gravity of the secondary, through
\begin{equation}
\eta=\left( \frac{M_{2}}{M_{1}} \frac{R_{\rm p}^3}{R_{2}^3} \right)^{1/2},
\end{equation}
where $M_1$ and $M_2$ are the primary and secondary mass, $R_{\rm p}$
is the periastron distance, and $R_2$ is the radius of the
secondary. With this definition, at a fixed mass ratio $q=M_2/M_1 \leq
1$, encounters with $\eta \gg 1$ have large impact parameters, while
$\eta \simeq 1$ marks collisions where the secondary is tidally
disrupted because the tidal field has become as intense as that
holding the star together. An important difference between encounters
with components of similar mass and those in which $q \ll 1$, as for
the disruption of stars by supermassive black holes in AGN
\citep{frank78,lacy82,carter83,rees88}, is that the periastron
distance is much smaller in the former, and in fact comparable to the
stellar radius. For a given intensity of the encounter, the periastron
distance can be written as
\begin{equation}
R_{\rm p}=R_2 \eta^{2/3} q^{-1/3}.
\end{equation}
Thus for a disruptive encounter with $\eta \simeq 1$, $R_{\rm p}
\simeq 200 R_2$ if $q=10^{-7}$, but $R_{\rm p} \simeq 2 R_2$ for
$q\simeq 1/3$. In the first case the disruption is a purely
gravitational encounter, while in the second it is additionally a
direct collision, which may modify the mass ratio substantially {\em
  during} the encounter.

The parameter $\eta$ can also be thought of as the ratio between the
dynamical time scale of the star (or its rotational break up period)
and the duration of the encounter. Thus for the intrinsic spin of the
secondary to be of any relevance, it must be rotating near the
Keplerian limit. The most rapidly spinning neutron stars detected in
LMXBs in our galaxy \citep{backer82} and in GCs \citep{hessels06} are
clearly below this threshold, by a factor 3-5, depending on the
assumed equation of state. It is thus reasonable to assume that in
most cases the neutron star spin is negligible for the purposes of the
encounters considered here, and we do so in what follows.

\subsection{Tidal capture and the formation of close binaries} 

When two compact stars pass close to each other so that $\eta \simeq
1$, the tidal perturbations produce accelerations of the stellar
material relative to the stellar center. The internal energy gained by
the stars is taken from the kinetic energy of their relative
motion. While the internal energy $\Delta E_{\rm T}$ gained by each
component is a small fraction of the star's internal energy, it can be
comparable with the kinetic energy of relative motion of the two
bodies at large separation. The two stars will thus become bound if
enough energy is absorbed in exciting the oscillations. The resulting
elliptical orbit immediatey after capture will have an initial
eccentricity only slightly less than unity. However, there will be
many successive passages at about the same periastron distance, and
these will lead to further energy transfer. Ultimately the orbit
should become nearly circular.

The value of $\Delta E_{\rm T}$ and thus the resultant condition for
tidal capture may be determined by requiring that the frequency of the
perturbation be slightly smaller than the natural frequencies of the
perturbed system. The tidal field excites non-radial modes of
oscillation in the secondary and, while there are many of these, those
most effectively excited have the smallest number of nodes, with an
angular frequency $\omega_{\rm osc}$ that is usually some two to three
times $\omega_{2} =(GM_2/R_2^3)^{1/2}$. Now, the angular frequency of
the encounter $\omega_{\rm p}$ is effectively $V_{\rm p}/R_{\rm p}$
and can be written as $\omega_{\rm p}\approx 2\omega_{2} (R_2/R_{\rm
  p})^{3/2}$. Thus for $R_{\rm p}/R_2=2$, $\omega_{\rm p}/\omega_{\rm osc} \approx
0.3$ for the lowest modes. However, for $R_{\rm
  p}/R_2=10$, $\omega_{\rm p}/\omega_{\rm osc}\approx 0.03$. For such
slow changes, the shape of each star will adjust to the changing form
of the equipotential surfaces and the net heating is markedly reduced.
The binaries formed by tidal capture are thus generally very hard.

Since the fraction of the initial angular momentum which is
transferred to stellar rotation can scarcely exceed a few percent
because of the relatively small stellar radius of gyration, one can
assume that the orbital angular momentum remains roughly constant.
Hence in a circular orbit of radius $R_{\rm c}$ and relative velocity
$V_{\rm c}$, $R_{\rm c}V_{\rm c}$ must equal the initial $R_{\rm
  p}V_{\rm p}$. With $V_{\rm p}$ obtained by the condition that
$(1/2)\mu_{\rm r}V^2_{\rm p}=GM_2^2/R_{\rm p}$, where $\mu_r=M_1
M_2/(M_1+M_2)$ and $V_{\rm c}$ computed from force balance in a
circular orbit, one finds $R_{\rm c} = 2R_{\rm p}$. To obtain more
exact results for the resulting conditions for tidal capture, the
excitation of the individual normal modes must be considered, and the
heating summed over all modes. Analysis of this effect by
\citet{press77} has taken the parabolic motion of the two stars
accurately into account, but in the tidal potential only terms varying
as $1/r^2$ and $1/r^3$ have been considered, an approximation valid
for $R_{\rm p}/R_2 \geq 3$. Detailed calculations were first carried
out using linear theory for $\Gamma=4/3$ \citep{press77, lee86} and
for $\Gamma=5/3$ \citep{lee86}. They were followed by many other
studies using both linear theory \citep{mcmillan87, koch92} and
numerical hydrodynamical calculations \citep{rass91}.  Such treatments
give $R_{\rm p}/ R_2$ between 2 and 3 as a condition for tidal capture
(the exact value depending on the properties of the stellar structure
model used). Thus an approximate condition for tidal capture may be
given as $R_{\rm p} \leq 3 R_2$ irrespective of the precise details
since for $R_{\rm p}/ R_2 \leq 3$, $\omega_{\rm p} \sim \omega_{\rm osc}$.

\begin{figure}
\includegraphics[width=\columnwidth,angle=-90,scale=1.]{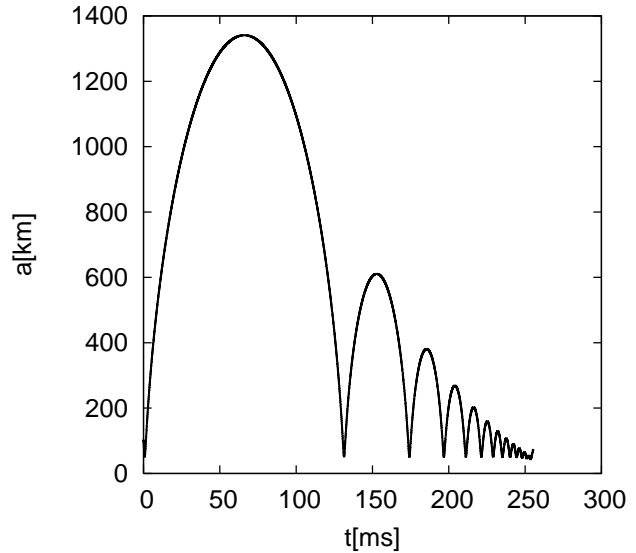}
\caption{The encounter of a neutron star with a black hole at
  relatively large impact parameter (run L$_3$) leads to a large
  number of periastron passages before tidal disruption. Here we show
  the separation (in km) between the neutron star and the black hole
  from the start of the simulation until full disruption of the star,
  taking about one quarter of a second.}
\label{fig:BHNSL3}
\end{figure}

For compact binaries, angular momentum losses to gravitational waves,
as well as dynamical mass ejection from the system must additionally
be considered.  The results for a black hole-neutron star encounter
with mass ratio $q=0.31$, $\eta=3$ (corresponding to $R_{\rm p}/ R_2 =
3.1$), where the neutron star is modeled as a polytrope with
$\Gamma=2$ (run L$_3$ in Table~\ref{tab:ICs}) are shown in
Figure~\ref{fig:BHNSL3}. As expected, the strong tides trigger complex
oscillations in the secondary star which can be clearly seen in the
variation of the maximum density in the core after the first
periastron passage. Further energy transfer occurs in subsequent
passages as the orbit becomes tighter and progressively circular,
until eventually some direct mass transfer takes place along with mass
stripping.  We find that the secondary is not shredded immediately,
but only after more than a dozen passages at about the same periastron
distance. Our condition for for tidal capture is in rough agreement
with $R_{\rm p}/ R_2 \leq 3$ although the inclusion of gravitational
waves as well as matter ejection from the system allows for a more
rapid variation of angular momentum disposal. For $\eta \simeq 1$, the
secondary star will experience a direct physical collision at least of
its outer layers. The process is rather complex with some of the gas
escaping entirely from the system on outbound trajectories. The core
may preserve its integrity for a few orbital periods around the
primary in a {\it common envelope} before finally coalescing. It is to
this problem that we now turn our attention.

\begin{deluxetable}{lcccccccc}
\tablecaption{Parameters for orbital encounters.\label{tab:ICs}} 
\tablewidth{0pt} 
\tablehead{\colhead{Run} & \colhead{Prim.} & \colhead{Sec.} & \colhead{$\Gamma$} 
& \colhead{$\frac{M_1}{M_{\odot}}$} & \colhead{$\frac{M_2}{M_{\odot}}$} 
& \colhead{$q=\frac{M_2}{M_1}$} & \colhead{$\eta$} & \colhead{$\frac{R_{\rm p}}{R_{2}}$} 
 }
\startdata 
L$_0$             & BH & NS & 2   & 4.51 & 1.4 & 0.31 & 1.0 & 1.5 \\
L$_0\Gamma_{5/3}$ & BH & NS & 5/3 & 4.51 & 1.4 & 0.31 & 1.0 & 1.5 \\
L$_1$             & BH & NS & 2   & 4.51 & 1.4 & 0.31 & 1.5 & 1.9 \\
L$_2$             & BH & NS & 2   & 4.51 & 1.4 & 0.31 & 2.0 & 2.3 \\
L$_2\Gamma_{5/3}$ & BH & NS & 5/3 & 4.51 & 1.4 & 0.31 & 2.0 & 2.3 \\
L$_3$             & BH & NS & 2   & 4.51 & 1.4 & 0.31 & 3.0 & 3.1 \\
NSNS              & NS & NS & 2   & 1.75 & 1.4 & 0.80 & 1.0 & 1.0 \\
BHWD              & BH & WD & 5/3 & 2.5  & 0.5 & 0.20 & 1.0 & 1.7 \\
\enddata
\end{deluxetable}

A summary of the most important aspects of the initial conditions thus
chosen is given in Table~\ref{tab:ICs}. There are six parabolic
encounters involving neutron stars with black holes, one double
neutron star collision, and one black hole--white dwarf encounter. The
parameter $\eta$ has been computed from the point mass orbital
parameters in Newtonian gravity. Naturally the centers of mass of each
star do not follow these solutions because of finite size effects and
the emission of gravitational waves, but they allow for a
characterization of each case. Note that since $\eta \propto R_{\rm
  p}^{3/2}$ and $L \propto R_{\rm p}^{1/2}$, the change in $\eta$ by a
factor 3 is actually a variation of $3^{1/3}\approx 1.4$ in orbital
angular momentum.

\subsection{Stellar disruption and disk formation}\label{sec:hydro}

For binary mergers, tidal disruption usually occurs after a single
periastron passage, because the system is already very tightly bound
by gravity. In the case of collisions, we find that the secondary is
not shredded immediately after the first interaction. It does,
however, lose a substantial amount of orbital energy and angular
momentum through two main channels: emission of gravitational waves
and transfer by gravitational torques through the formation of the
tidal bulge. The bulge in effect deforms the spherical star into a
bar, which the primary can then spin-up during the brief
encounter. The strong tides additionally trigger complex, nearly
radial oscillations in the secondary which can be clearly seen in the
variation of the maximum density in the core after periastron
passage. Some direct mass transfer occurs as well, along with mass
stripping which can form a first accretion structure around the
primary, which is more massive for low initial impact
parameters. Additionally, some matter is flung out through the
exterior Lagrange point to large distances, creating a tidal tail.

\begin{figure}
\includegraphics[width=\columnwidth,angle=0,scale=1.]{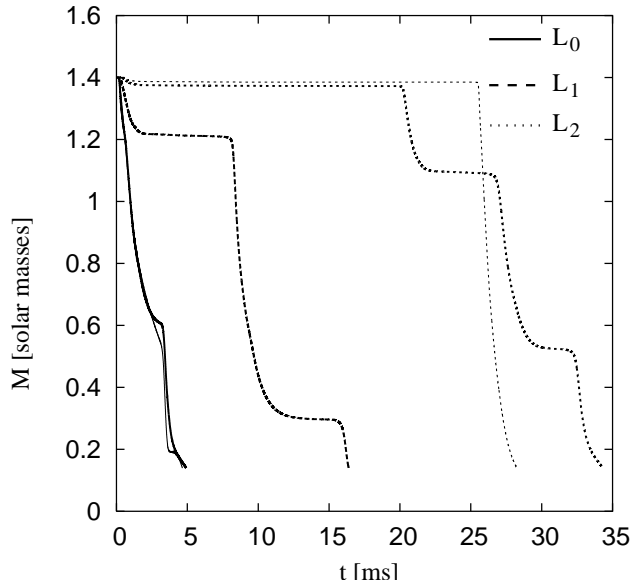}
\caption{The neutron star core mass decreases during the collision
  with a black hole by tidal mass stripping. As the strength of the
  encounter, $\eta$, decreases, the neutron star survives for a
  greater number of periastron passages before being eventually
  shredded. The secondary's mass (in solar masses) is shown for runs
  L$_{0}$, L$_0\Gamma_{5/3}$, L$_{1}$, L$_{2}$ and L$_2\Gamma_{5/3}$
  until disruption. The thick (thin) lines are for $\Gamma=2$
  ($\Gamma=5/3$).}
\label{fig:BHNScoremass}
\end{figure}

\begin{figure}
\includegraphics[width=\columnwidth,angle=0,scale=1.]{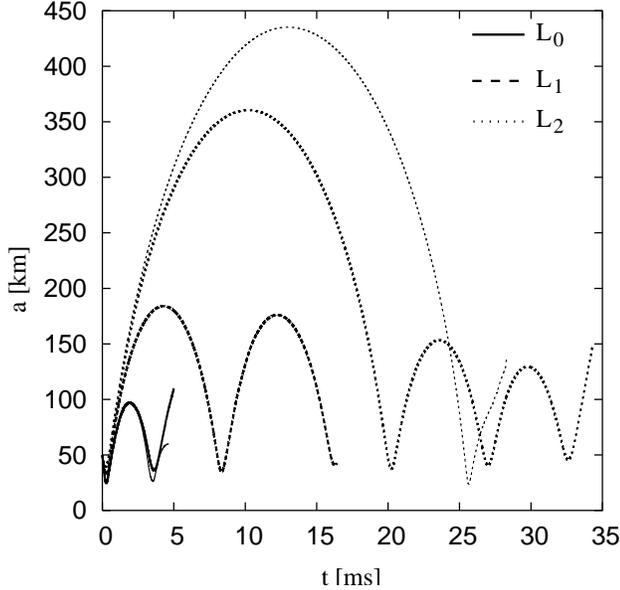}
\caption{The separation (in km) between the neutron star and the black
  hole is shown until disruption of the neutron star for runs L$_{0}$,
  L$_{0}\Gamma_{5/3}$, L$_{1}$, L$_{2}$ and L$_{2}\Gamma_{5/3}$. Note
  the change in scale on the time axis when compared with
  Figure~\ref{fig:BHNSL3}. The thick (thin) lines are for $\Gamma=2$
  ($\Gamma=5/3$).}
\label{fig:BHNSsep}
\end{figure}

\begin{figure}
\includegraphics[width=\columnwidth,angle=0,scale=1.]{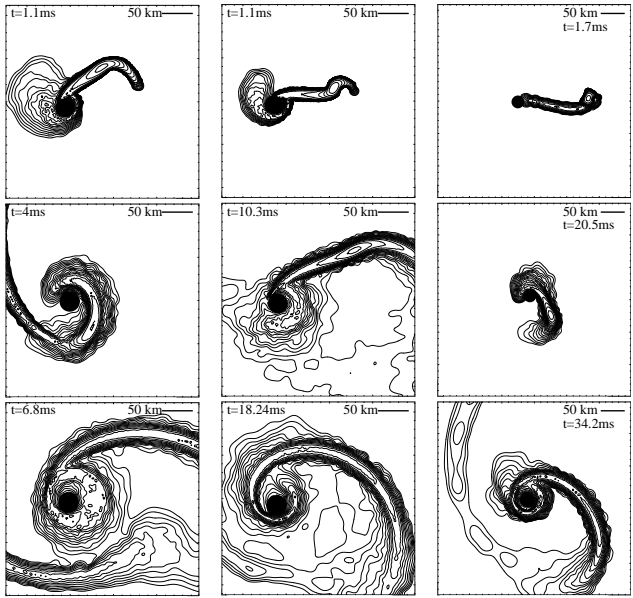}
\caption{Parabolic collisions of neutron stars with black holes. Logarithmic contours of density in the orbital plane (equally
spaced every 0.25 dex) are shown for runs L$_{0}$ (left), L$_{1}$
(middle) and L$_{2}$ (right). The lowest contour in bold is at $\log
\rho \; [\mbox{g~cm$^{-3}$]}=10$. The time elapsed since the beginning
of the simlulation is indicated in each panel (increasing top to
bottom), as is the distance scale. Note the different number of
periastron passages and tidal tails formed until final disruption in
each case.}
\label{fig:BHNSL0L1L2}
\end{figure}

In runs L$_0$ and L$_1$, the neutron star core (which now contains
$\simeq 1~M_{\odot}$ and $1.2~M_{\odot}$ respectively, see
Figure~\ref{fig:BHNScoremass}) does not survive the second encounter,
and forms a massive disk around the black hole, as well as an
elongated tidal tail. In run L$_2$ there is enough energy and angular
momentum at the outset that the core is able to return a third time,
after which it too is fully disrupted and forms a disk (see
Figure~\ref{fig:BHNSsep}). Each successive passage feeds the accretion
disk and simultaneously forms a tidal tail, which is not set to
collide with previous ejections. We show in
Figure~\ref{fig:BHNSL0L1L2} the time evolution for different initial
impact parameters, projected onto the orbital plane. Even with three
periastron passages, the collision and disruption are essentially over
after $\simeq 50$~ms, because the initial passage drains enough
orbital energy and angular momentum to bind the system very
effectively. 

For runs carried out with a soft equation of state, $\Gamma=5/3$, the
results are qualitatively the same for low impact parameter, with the
main difference being in the spatial extent of the accretion disk
formed, and in that of the tidal tails, both being greater than for
$\Gamma=2$. For large impact parameter, the gravitational interaction
is quantitatively different during the first periastron passage, since
the star effectively resembles a point mass to a greater degree in run
L$_2\Gamma_{5/3}$. Angular momentum and energy transfer through
torques is thus less efficient and the core of the neutron star is
transferred to a higher orbit than in run L$_{2}$ (note the difference
in secondary apocenter values in Figure~\ref{fig:BHNSsep}). At the
secondary passage $\simeq 26$~ms after the start of the simulation,
however, the star is fully shredded by tidal forces and the final disk
forms promptly.

For the disruption of a star by a {\em supermassive} black hole, the
former moves essentially in the fixed background metric imposed by the
hole, and this allows for a simplified treatment of the dynamics. It
was found in earlier studies \citep{rees88}, both through analytical
considerations and direct numerical simulation, that essentially half
the mass of the star is dynamically ejected, while the remaining half
is captured by the black hole, being on eccentric trajectories which,
with variable delay, will bring them back to the vicinity of the
primary.

\begin{figure}
\includegraphics[width=\columnwidth,scale=1.]{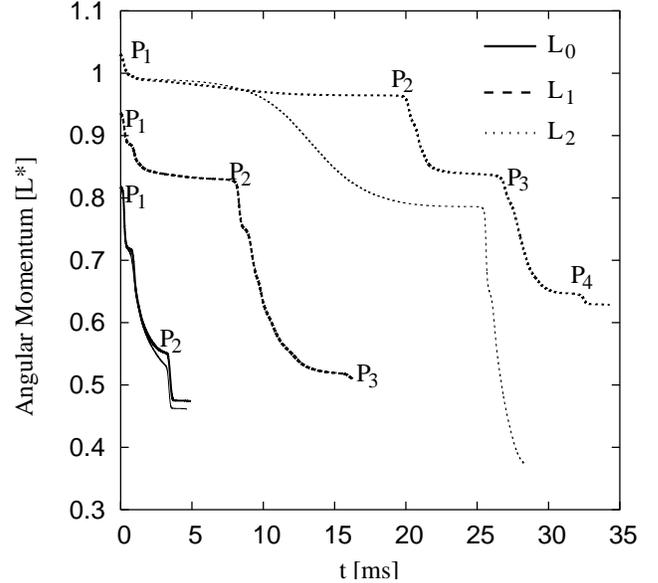}
\caption{The total angular momentum as a function of time is shown for
  runs L$_{0}$, L$_{0}\Gamma_{5/3}$, L$_{1}$, L$_2$ and
  L$_{2}\Gamma_{5/3}$, in units of $L^{*}=M_{\rm tot}\sqrt{G \mu
    R_{\rm NS}}$, where $M_{\rm tot}$ is the total mass and $\mu$ is
  the reduced mass. For each case, successive periastron passages are
  marked on the curve as P$_{1}$, P$_{2}$, ... .The thick (thin) lines
  are for $\Gamma=2$ ($\Gamma=5/3$).}
\label{fig:BHNSangmom}
\end{figure}

There are two important differences between such a scenario and that
considered in the present study. First, as already noted, the mass
ratio is of of order 0.1-1 instead of 10$^{-7}$, which can produce a
direct collision with accompanying mass transfer simply by virtue of
the small periastron distance of a disruptive event. Second, at the
small distance scales implied by the fact that we are considering
compact stellar mass objects, gravitational wave emission is intense,
and can drain a substantial fraction of the total kinetic energy and
angular momentum during a single passage (see
Figure~\ref{fig:BHNSangmom}). Thus the system becomes non-conservative
from a point of view of orbital dynamics, and the previous reasoning
does not apply.

After the neutron star has been fully disrupted, the remnant consists
of a black hole surrounded by a torus, and a series of tidal tails,
depending on the number of periastron passages which occurred. The
tori are typically 200-300~km across and contain $M_{\rm disk}\simeq
0.1 M_{\odot}$. They are comparable in size and mass to those
encountered during binary mergers, with densities $\rho \simeq
10^{11}-10^{12}$~g~cm$^{-3}$ and internal energies $u \simeq
10^{18}-10^{19}$erg~g$^{-1}$, equivalent to 1-10~MeV/baryon.  The
binary interaction is violent and complex, though, and this can be
qualitatively seen in the fact that by the end of our calculations
(roughly 20~ms) the accretion structures still show significant
deviations from azimuthal symmetry.

\begin{figure}
\includegraphics[width=\columnwidth,angle=0,scale=1.]{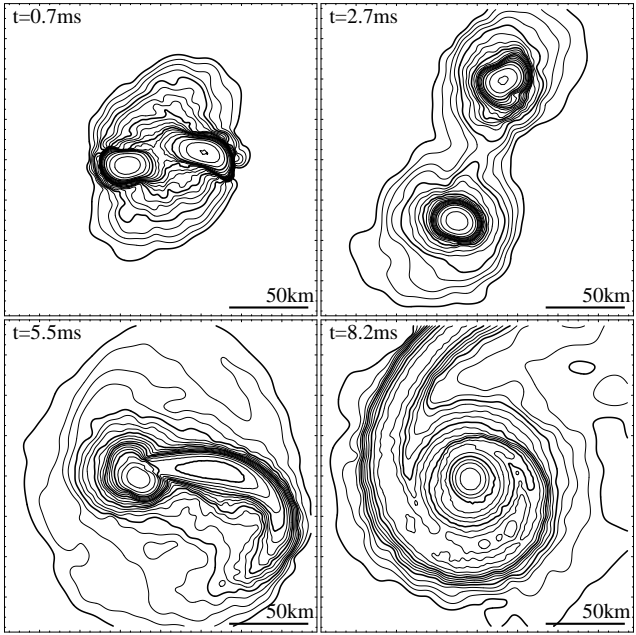}
\caption{Logarithmic contours of density in the orbital plane (equally
  spaced every 0.25 dex) for the collision of two neutron stars (run
  NSNS) on a parabolic orbit with initial mass ratio $q=0.8$. The
  lowest contour in bold is at $\log \rho \;
  [\mbox{g~cm$^{-3}$]}=10$. After an initial passage the secondary is
  entirely disrupted and wraps around the primary, forming an envelope
  and a single tidal tail.}
\label{fig:NSNScoll}
\end{figure}

For double neutron star encounters, the mass ratio is even greater (we
have computed one collision for $q=0.8$, which is likely to be a lower
bound for such systems). For the case considered with adiabatic index
$\Gamma=2$, the two stars are actually the same size, and directly
impact each other for an encounter strength $\eta=1$.  The initial
collision binds the stars in an elliptical orbit but does not lead to
the formation of a significant tidal tail. The less massive secondary
is strongly distorted and spun up, and a bridge of material
temporarily joins the two stars. Upon a second passage the secondary
is entirely shredded and wraps around the primary (see
Figure~\ref{fig:NSNScoll}). The material from each stars remains
largely separate, with that from the primary remaining essentially in
the core, in a manner similar to what is obtained for mergers of
unequal mass neutron stars.

\begin{figure}
\includegraphics[width=\columnwidth,angle=0,scale=1.]{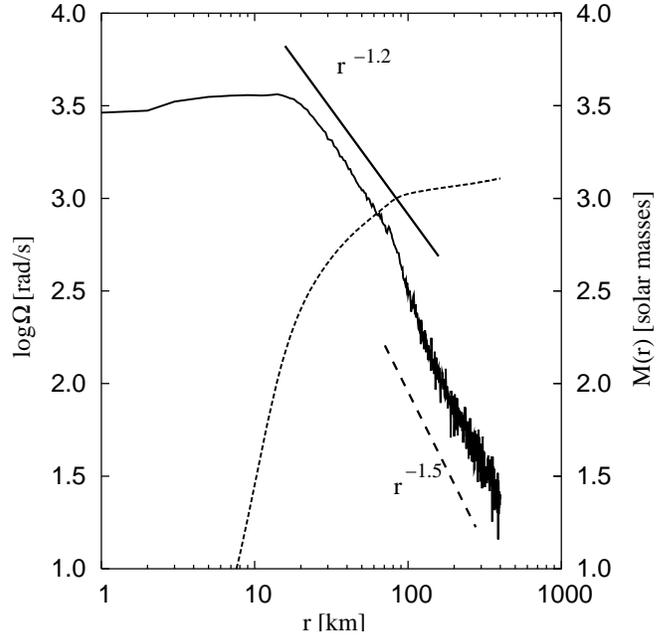}
\caption{Azimuthally averaged angular velocity profile (solid line)
and enclosed mass (dashed line) in the double neutron star collision
remnant. The rapidly rotating inner core (P=2.1~ms) is surrounded by a
large envelope. The rotation of the tidal tail is visible at radii
greater than 100~km, and reference power laws are given for the
envelope and tail.}
\label{fig:NSNSrot}
\end{figure}

The final remnant in this case consists of a slightly differentially
rotating core of 2.4$M_{\odot}$ with radius $R_{\rm core} \simeq
20$~km and a maximal rotation frequency of $\Omega_{\rm max} \simeq
3000$~rad/s, corresponding to a period of 2.1~ms (see
Figure~\ref{fig:NSNSrot}). This is surrounded by an envelope out to
$\sim$~80~km containing $M_{\rm env}\simeq 0.49 M_{\odot}$, where the
orbital frequency is $\Omega \propto r^{-1.2}$. In the outer tail of
material the rotation is essentially Keplerian, with $\Omega \propto
r^{-1.5}$ (the rotation profile is shown as well in
Figure~\ref{fig:NSNSrot}). Note that the core is above the threshold
for collapse of a cold, non--rotating configuration in most equations
of state, but could avoid this given the rapid and differential
rotation present \citep{cook94,baumgarte00}. In addition significant
heating of the core can raise the critical threshold mass for collapse
to $~\simeq 1.35 M_{\rm cold}$ \citep{shibata06}. In this particular
case the core could conceivably remain stable for a longer time,
spinning down on a secular time scale due to the emission of
gravitational waves or magnetic torques. If it were to indeed form a
black hole, the distribution of angular momentum in the envelope is
such that only material between 50 and 100~km, amounting to
0.17$M_{\odot}$, would have enough centrifugal support to form an
accretion disk.

For completeness we have also considered the interaction of a low mass
($M_{\rm WD}=0.5M_{\odot}$) white dwarf with a compact object ($M_{\rm
  co}=2 M_{\odot}$). The latter could be either a massive neutron star
or a low mass black hole. Numerically it is irrelevant because the
characteristic scales of the two objects are so different that it is
impossible to resolve simultaneously the black hole horizon (or
neutron star surface) and the entire white dwarf. For the actual
calculation the absorbing accretion boundary has been placed
100~gravitational radii from the center of the compact object.

\begin{figure}
\includegraphics[width=\columnwidth,angle=0,scale=1.]{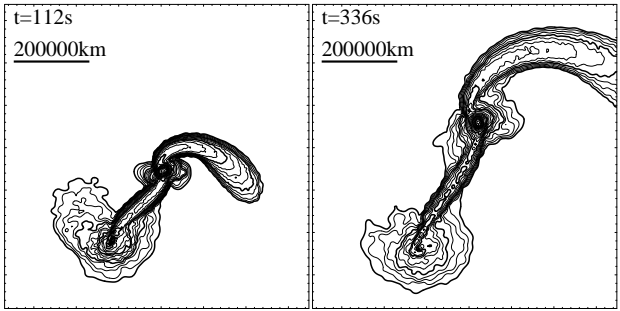}
\caption{Logarithmic density contours (equally spaced every 0.25 dex)
for the encounter between a low mass white dwarf and a compact object.
The lowest contour in bold is at $\log \rho \;
[\mbox{g~cm$^{-3}$]}=3$. Note the difference in scales (spatial and
temporal) when compared to those in Figures~\ref{fig:BHNSL0L1L2} and
\ref{fig:NSNScoll}. The higher compressibility of the material also
makes for a wider distribution of ejected material and a generally
wider distribution of gas.}
\label{fig:WDBHcoll}
\end{figure}

The white dwarf is modeled as a polytrope with adiabatic index
$\Gamma=5/3$, appropriate for a cold, non-relativistic degenerate
configuration, and the pressure is thus given by $P=K_{\rm n.r.}
\rho^{5/3}$, where $K_{\rm n.r.}=(3/\pi)^{2/3}h^2/(20 m_e
(2m_p)^{5/3})$. With the given mass, the stellar radius is $R_{\rm
WD}=1.1 \times 10^{9}$~cm, a factor $f \simeq 800$ larger than our
standard neutron star. In fact with the same input physics one could
simply scale the results from the BH-NS interaction (given an
identical mass ratio) by increasing distances by a factor $f$ and
temporal scales by $f^{3/2}\simeq 2.3 \times 10^{4}$ (giving hundreds
of seconds instead of tens of milliseconds). This is strictly not
correct, however, since gravitational radiation reaction introduces an
absolute scale into the problem, and energy and angular momentum
losses through this channel are insignificant in comparison to the
BH-NS case. The interaction proceeds then at a more leisurely pace, so
much so that we were unable to follow it to a second periastron
passage, even though the simulation covered nearly one thousand
seconds. At this stage stripping during the close passage has formed a
torus around the compact object, linked to the stellar core by a long
and narrow bridge of material (see Figure~\ref{fig:WDBHcoll}). The
core itself is rapidly spinning due to gravitational torques exerted
by the primary, and the typical tidal tail has formed at large radii.

Since we are considering encounters in which disruption occurs
practically by construction (given the choice of the parameter
$\eta$), the gravitational torques exerted on the secondary are of
comparable magnitude in all cases (with appropriate scalings). We find
that the angular frequency of the core after the first periastron
passage is $\simeq (0.2-0.4) \Omega_{0}$, where
$\Omega_{0}=(GM_2/R_2^{3})^{1/2}$ is the break up rotation frequency
of the unperturbed secondary. Thus the spin periods are $\simeq 2$~ms
and $\simeq 8$~s for encounters involving neutron stars and white
dwarfs respectively.

\begin{deluxetable}{lccc}
\tablecaption{Disk, tail and ejected masses.\label{tab:masses}}
\tablewidth{0pt} \tablehead{\colhead{Run} & \colhead{$M_{\rm
disk}/M_{\odot}$} & \colhead{$M_{\rm tail}/M_{\odot}$} &
\colhead{$M_{\rm ej.}/M_{\odot}$} } 
\startdata 
L$_0$ & 0.23 & 0.15 & 0.05 \\ 
L$_{0}\Gamma_{5/3}$ & 0.25 & 0.11 & 0.04 \\
L$_1$ & 0.16 & 0.31 & 0.13 \\ 
L$_2$ & 0.09 & 0.47 & 0.19 \\
L$_{2}\Gamma_{5/3}$ & 0.11 & 0.19 & 0.04 \\
NSNS   & 0.17 & 0.13 & 0.03 \\
BHWD   & 0.13 & 0.35 & 0.24 \\
\enddata
\end{deluxetable}

The mass of the disk (see Table~\ref{tab:masses}) present at the end
of the calculation as a result of the encounter is computed from the
fluid elements in close, essentially circular orbit about the central
object. It is typically $M_{\rm disk} \simeq (0.1-0.3) M_{\odot}$,
regardless of the type of encounter. Likewise, the mass of the tidal
tails, $M_{\rm tail}$, is obtained by adding over all the fluid
elements within these structures, whether they are bound to the
central mass or not. For black hole-neutron star encounters there is a
clear trend of decreasing disk mass and increasing tail mass as the
initial impact parameter grows at fixed compressibility. The total
amount of matter dynamically ejected (computed as that with positive
total energy) also increases, and is a result of the greater number of
ejection episodes associated with larger values of the orbital angular
momentum. For run L$_{2}$ we find that nearly 0.2$M_{\odot}$ are thus
lost. For the double neutron star collision the characteristics of the
system (total mass and mass ratio) are such that the ejected mass is
significantly lower, although a substantial tail is still present. We
will return to the implications of mass ejection below. Including the
effects of General Relativity will likely alter these values, by up to
one order of magnitude if the differences between calculations
performed in Newtonian theory and those using GR for mergers are used
as guidance.

\subsection{Tidal tails and mass ejection}\label{sec:tails}

For the two members of the system to come together and eventually
merge, they must lose energy and angular momentum. This can be
achieved through the emission of gravitational waves, or the ejection
of matter, or a combination of both. In either a binary system in
circular orbit or a parabolic approach there is a substantial amount
of rotation. Since a small amount of matter, removed to a large
radius, can carry a great deal of angular momentum, the formation of
tidal tails out of material stripped through the outer Lagrange points
in the effective potential is an efficient way to produce a single
object at the center. This is the fundamental reason why such
structures form in either stellar or galactic collisions.

\begin{figure}
\includegraphics[width=\columnwidth,angle=0,scale=0.85]{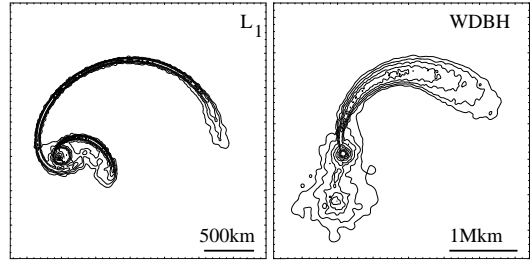}
\caption{Tidal tails formed by the disruption of a neutron star by a
  black hole (run L$_{1}$) and a white dwarf (run WDBH). For the
  neutron star disruption the impact parameter was such that two
  periastron passages occurred, giving rise to two distinct ejection
  structures thousands of kilometers across by the end of the
  simulation. For the case involving the white dwarf, only one passage
  has occurred and the core of the star is still present. Note the
  much larger spread in the distribution of the fluid, due to the
  lower adiabatic index used in the case of the white dwarf.}
\label{fig:tails}
\end{figure}

Just as the vanishing orbital energy of a parabolic orbit implies
successive periastron passages as the core becomes more and more bound
to the primary, so too it is easier to dynamically unbind matter to
infinity when compared to a bound binary. As each passage proceeds a
new ejection episode occurs, giving rise to an additional tail
(Figure~\ref{fig:tails}). The amount of mass thus ejected, and shown
in Table~\ref{tab:masses}, is considerably greater than for a binary
coalescence, by about one order of magnitude \citep{lee01}.

In all cases when multiple ejection events occur (we observed up to
three for run L$_{2}$), the velocities and orientations are such that
the initial tail will not be overtaken by latter ones. This is simply
because the first one has a clear head start, but also because
subsequent events occur at different orbital phases.

An interesting point regarding the tails concerns their
hydrodynamic stability. Their motion is essentially ballistic,
dominated by the potential well of the central mass. They are
nevertheless susceptible to the varicose, or sausage instability first
identified by Rayleigh in 1899 \citep{chandra61}. This is due to
self-gravity and occurs for cylindrical configurations of an
incompressible fluid for perturbations with wavelength $\lambda >
\lambda^{*}=2 \pi R_{\rm cyl}/x^{*}$, where $x^{*}\simeq 1$ and
$R_{\rm cyl}$ is the radius of the cylinder. The fastest growing mode
has $x=0.58$, wavelength $\lambda\simeq 11 R_{\rm cyl}$ and a growth
time $\tau=4/(4 \pi G \rho)^{1/2}$ \citep{chandra61}. For sufficiently
stiff equations of state (with adiabatic index $\Gamma > 2.5$), even
though not strictly incompressible, this is actually seen in numerical
simulations \citep{rasio94,lee00}: on a time scale given approximately
by $\tau$, condensations form, and are separated roughly by the
wavelength of the fastest growing mode given above. For softer
equations of state (even such as the one used here for neutron stars
when $\Gamma=2$) the effect is not present, and even less so for the
calculation involving the neutron star with $\Gamma=5/3$ and the white
dwarf (Figure~\ref{fig:tails}).

\section{Emission of gravitational waves}\label{sec:gwaves}

In the case of merging binaries, the early gravitational waveforms,
when the separation is much larger than the stellar radius, can be
computed using the weak field approximation analytically, and from
them the stellar masses may be accurately determined. As the stars
become distorted by the tidal field, the signal deviates form this
solution and finite size effects accelerate the decay. The secondary
is then fully accreted by the black hole (if one is present) or
tidally disrupted, and the emission abruptly ceases. The precise
frequency where this occurs can lead in principle to accurate
determinations of the neutron star radius \citep{faber02}, and thus,
since the mass is already known, to useful constraints on the equation
of state at supra-nuclear densities.

\begin{figure}
\includegraphics[width=\columnwidth,angle=0,scale=1.]{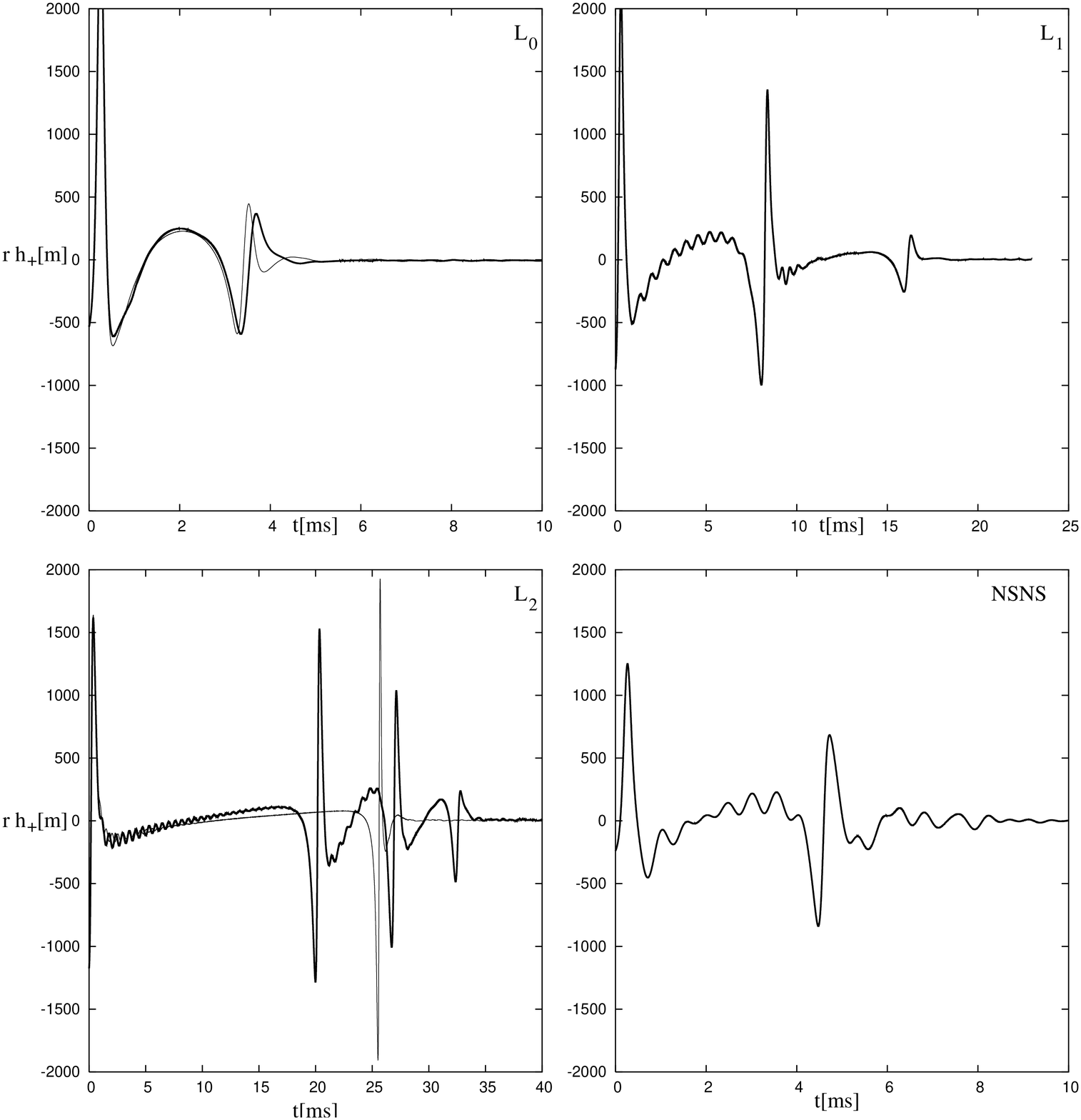}
\caption{Gravitational waves emitted as a function of time for runs
  L$_{0}$, L$_0\Gamma_{5/3}$, L$_{1}$, L$_{2}$, L$_2\Gamma_{5/3}$, and
  NSNS as seen by an observer placed along the rotation axis ($r
  h_{+}$ is plotted (in m), where $r$ is the distance from the source
  to the observer). The thick (thin) lines are for $\Gamma=2$
  ($\Gamma=5/3$). The vertical range is identical in all frames, but
  note the different scaling on the time axis. In all runs except for
  L$_{0}$, L$_{0}\Gamma_{5/3}$, and L$_{2}\Gamma_{5/3}$, the
  oscillations of the neutron star core following the first and even
  second periastron passage are clearly visible. When varying the
  stiffness of the equation of state, the waveforms are nearly
  indistinguishable for small impact parameter and clearly separated
  for large values, due to the difference in the associated orbital
  evolution.}
\label{fig:BHNSgw}
\end{figure}

The total mass and relative velocities involved in a parabolic
encounter are similar to those encountered in close binaries, and so
we would expect the characteristic frequencies and strength, or
amplitude of the signal in gravitational waves, to be comparable in
this case. The biggest difference arises, and is crucial in terms of
detectability, because the collision does not involve a leisurely
spiral in over many orbital periods, and as thus lacks the
characteristic precursor chirp signal which would slowly sweep through
the interferometer's detection band. Figure~\ref{fig:BHNSgw} shows the
computed waveforms (one polarization is given) for black hole-neutron
star encounters and the double neutron star collision.

A neutron star ($M_2$) approaching a stellar mass black hole ($M_1$)
with $\eta \sim 1$ will be disrupted in a single passage and the
particles in the disrupted remnant follow approximately independent
Keplerian orbits. The detectable gravitational signal will thus have a
burst-like behaviour, roughly characterized by an amplitude
\begin{multline}
  h \sim{G M R_1 \over c^2D R_{\rm p}} 
  \sim 10^{-22} \; 
  \eta ^{-2/3}
  \left({D \over 100 \, {\rm Mpc}}\right)^{-1}
  \times \\
  \left({M_1 \over 10 \, M_\sun}\right)^{2/3}
  \left({R_2 \over 10 \, {\rm km}}\right)^{-1} 
  \left({M_2 \over 1.4 \, M_\sun}\right)^{4/3},
\end{multline}
and frequency
\begin{multline}
  f \sim \left({G M_1 \over R_p^3}\right)^{1/2}
  \sim 1.4 \times 10^{4} \, {\rm Hz} \;
  \times \\
  \eta^{-1}
  \left({R_2 \over 10 \, {\rm km}}\right)^{-3/2}
  \left({M_2 \over 1.4 \, M_\sun}\right)^{1/2}.
\end{multline}
Here $M$ denotes the total mass, $M_{1}+M_{2}$. LIGO will be able to
detect gravitational wave from impact involving neutron stars and
stellar mass black holes if $\eta \leq 1$ and the distance is $D\leq
50$ Mpc. In what follows, we compare these simple estimates of
gravitational radiation against more detailed results obtained with
our numerical scheme.

The signal exhibits a local peak at each periastron passage,
until the time when the star is completely disrupted and the amplitude
vanishes (note the different time scales on each plot in
Figure~\ref{fig:BHNSgw}). In runs L$_{1}$, L$_{2}$ and NSNS, smaller
oscillations of decaying amplitude after each passage (but perhaps the
last) are also clearly present when $\Gamma=2$. For instance, for run
L$_{2}$ between 3 and 15~ms, their frequency is $\nu_{\rm osc} \simeq
1750$~Hz. This is {\em not} due to the rotation of the neutron star
(the spin frequency of the core is at this point approximately
$\nu_{\rm spin} \simeq 320$~Hz) but essentially to radial vibration
modes excited by the action of the primary at periastron passage. The
frequency of such modes is close to the natural value given by
hydrostatic equilibrium, namely $ \nu_{\rm osc} \simeq (G M /
R^{3})^{1/2}/2 \pi \simeq 1400$~Hz. The small impact parameter and
rapid disruption of the star in run L$_{0}$, and the high
compressibility used in run L$_2\Gamma_{5/3}$ does not allow for a
clear manifestation of such oscillations.

\begin{figure}
\includegraphics[width=\columnwidth,angle=0,scale=1.]{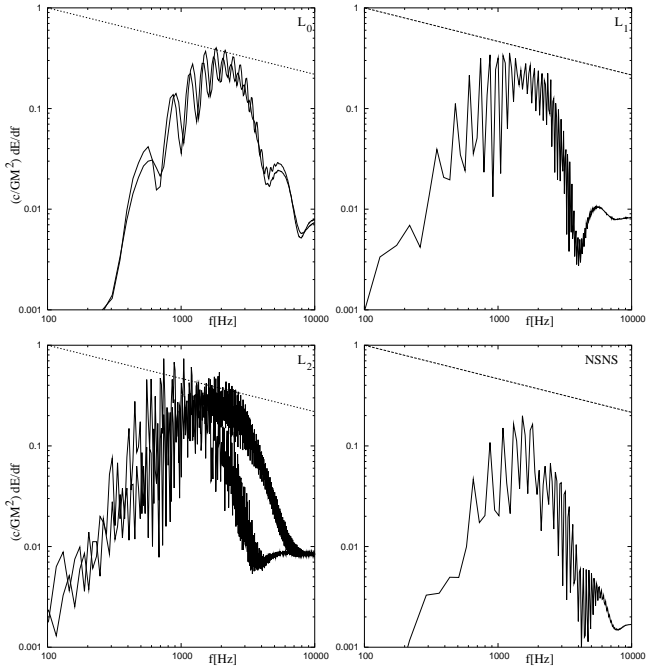}
\caption{Gravitational wave energy spectrum for runs L$_{0}$,
  L$_{0}\Gamma_{5/3}$, L$_{1}$, L$_{2}$, L$_{2}\Gamma_{5/3}$ and
  NSNS. The ranges are identical in all frames. The reference power
  law is $dE/df \propto f^{-1/3}$, the characteristic spectrum for the
  in-spiral of a point mass binary. The thick (thin) lines are for
  $\Gamma=2$ ($\Gamma=5/3$). The oscillating nature of these spectra
  is due to the finite extent of wave trains present in the amplitudes
  during different segments of temporal evolution.}
\label{fig:BHNSgwsp}
\end{figure}

\begin{figure}
\includegraphics[width=\columnwidth,angle=0,scale=1.]{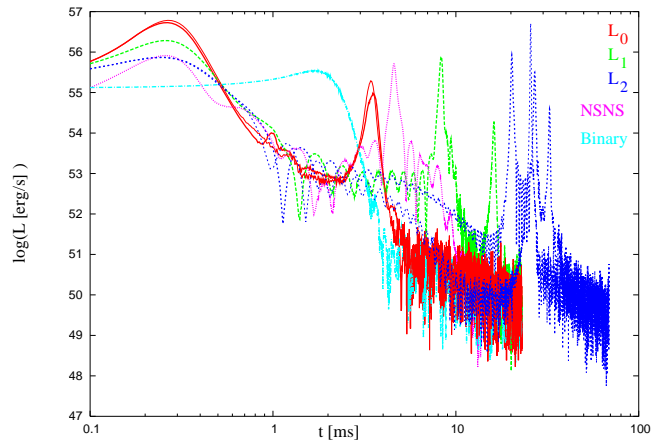}
\caption{Power radiated in gravitational waves for black hole-neutron
  star encounters and the double neutron star collision. There is one
  successively weaker peak for each periastron passage until complete
  tidal disruption. The curve labeled ``Binary'' shows the luminosity
  computed for a black hole-neutron star binary with the same mass
  ratio, $q=0.31$ and initial separation, $r_{i}=3.7 R_{\rm NS}$, as
  runs L$_0$ through L$_{2}$, for which disruption occurs promptly
  after the beginning of mass transfer \citep{lee01}. The thick (thin)
  lines are for $\Gamma=2$ ($\Gamma=5/3$). When varying the stiffness
  of the equation of state, the luminosities are nearly
  indistinguishable for small impact parameter and clearly separated
  for large values, due to the difference in the associated orbital
  evolution.}
\label{fig:BHNSGWlum}
\end{figure}

The power spectrum of these gravitational waves is shown in
Figure~\ref{fig:BHNSgwsp}. There are peaks at 1900, 1600 and 1000~Hz
for runs L$_{0}$, L$_{1}$ and L$_{2}$, respectively. These correspond
to the interaction time for each run, which increases (thus decreasing
the frequency) as the impact parameter (or equivalently, the total
angular momentum in the system) increases. Secondary periastron
passages make this peak wider at higher frequencies. This is
particularly evident when comparing the results of runs L$_2$ and
L$_2\Gamma_{5/3}$. The latter has a single late-time periastron
passage (instead of three for the former), and the peak is broadened
to $\simeq 2000-3000$~Hz. We note that the oscillation frequency of
the core, visible in the signal when $\Gamma=2$, is undetectable in
these spectra, being buried within the main peak due to the binary
interaction (this is to be expected, since by definition of a tidally
disruptive event, the interaction time is roughly equal to the free
fall time scale of the star). The spectrum for run NSNS shows a
similar overall structure. We note that it is difficult to extract a
clean spectral signal with a high signal to noise ratio, because the
simulated time is quite limited. Secondary high frequency variations
at $\simeq 5000$~Hz can be seen in the spectra for runs L$_{0}$,
L$_{1}$ and NSNS and are related to the asymmetry of the waveform
around the time of closest approach. The large number of periastron
passages in run L$_{2}$ makes the spectrum more noisy and this feature
is not observed. Finally, we also show the gravitational wave
luminosity in Figure~\ref{fig:BHNSGWlum}, the integration of which
reveals that the efficiency for their emission is $\epsilon= E_{\rm
  GW} /(M c^{2})= 10^{-2}, 7 \times 10^{-3}, 6.8 \times 10^{-3}, 5.9
\times 10^{-3}$ for runs L$_{0}$, L$_{1}$, L$_{2}$ and NSNS
respectively. For comparison, the efficiency in the case of a merging
binary with the same mass ratio is $\approx 5\times 10^{-3}$
\citep{lee01}. We note also from Figure~\ref{fig:BHNSGWlum} that the
maximum power radiated during a collision and even during subsequent
close approaches can be significantly higher than that in a merging
binary with the same mass ratio, due to the larger velocities in an
eccentric orbit at periastron.

\section{The rates of collisions and tidal captures of compact objects
  in globular clusters}
\label{sec:rates}

The highest stellar densities in GCs are reached in the core, in
particular during core collapse. In this section we estimate the
collision and tidal capture rate as a function of time by first
reconstructing the core evolution of a typical post-core-collapse GC.
We then re-scale the resulting encounter rate evolution with the
measured mass and size distribution of GCs in galaxies to obtain the
rate per host galaxy. Finally, for a given galaxy luminosity density
distribution, we obtain per volume of space and as function of time,
the expected global rate of collisions and tidal capture of compact
objects in the cores of GCs. The details of the encounter rate
calculation will be presented in an forthcoming paper (van de Ven et
al., in prep.), of which we give here a summary focussing on close
encounters between compact stars.

\subsection{Close encounters}
\label{sec:encounters}

We assume the different types of stellar objects, $i$, are distributed
homogeneously within a spherical core of radius $\rc$, each with
fractional number $f_i \le 1$ and total number density $\nc$.  We
further assume that the stars follow a Maxwellian velocity
distribution function with dispersion $\sigc$. Together with the
dominating gravitational focusing, this means we can approximate
\citep{heggie75} the total collision rate as
\begin{multline}
  \label{eq:nucoll}
  \nucol = 
  2.1 \times 10^{-3} \; \mathrm{Gyr}^{-1} \; 
  \frac{f_1\,f_2}{1+\delta_{12}}
  \times \\
  \Bigl(\,\frac{n_c}{10^6\;\mathrm{pc}^{-3}}\,\Bigr)^2
  \Bigl(\,\frac{r_c}{0.1\;\mathrm{pc}}\,\Bigr)^3
  \Bigl(\,\frac{\sigma_c}{10\;\mathrm{km\,s}^{-1}}\,\Bigr)^{-1}
  \times \\
  \Bigl(\,\frac{M_1+M_2}{1\;\mathrm{M}_\odot}\,\Bigr)
  \Bigl(\,\frac{\Rmin}{10\;\mathrm{km}}\,\Bigr),
\end{multline}
where $\delta_{12}=1$ if type 1 and 2 are equal, and $\delta_{12}=0$
otherwise. The stars have masses $M_i$, and their separation at
closest approach is given by $\Rmin$, which we take to be the sum of
the stellar radii, i.e., $\Rmin = R_1 + R_2$.

In addition, we also consider encounters in which stars pass close
enough to each other to form a binary by transfering orbital energy to
internal stellar oscillations. We adopt a cross section for tidal
capture of the form \citep{lee86}
\begin{equation}
  \label{eq:Sigtid}
  \Sigma_\mathrm{tid} = 
  a \left( \frac{v_\mathrm{inf}}{v_{\star,2}} \right)^{-\beta}
  R_2^2,
\end{equation}
where $v_\mathrm{inf}$ is the relative velocity at infinity, and
$v_{\star,2} = (2GM_2/R_2)^{1/2}$ is the escape velocity at the
surface of the secondary, captured star. We use the fitting functions
of \cite{kimlee99} to obtain the amplitude $a$ for encounters between
different stellar types, while the slope $\beta \simeq 2.2$ in all
cases. In case of collisions dominated by gravitational focussing, the
cross section is of the same form, but with slope $\beta = 2$.
Henceforth, we can express the tidal capture rate $\nutid$ in the same
way as the collision rate in equation~\eqref{eq:nucoll}, but with
closest approach given by
\begin{equation}
  \label{eq:Rmintid}
  \Rmin = 
  \Gamma(2-\beta/2) \frac{a}{\pi} \frac{M_2}{M_1+M_2} 
  \left( \frac{2\sigc}{v_{\star,2}} \right)^{2-\beta} R_2,
\end{equation}
where $\Gamma$ is the complete gamma function.

\begin{deluxetable}{*{7}{r}}
  \tablecaption{Relative encounter rates. \label{tab:rates}}
  \tablewidth{0pt}
  \tablehead{ \colhead{type~1} & \colhead{type~2} &
    \colhead{$M_2/\mathrm{M}_\odot$} & \colhead{$R_2/\mathrm{km}$} &
    \colhead{$\xicol$} & \colhead{$\nutid/\nucol$} &
    \colhead{$\Rmin/R_2$} }
  \startdata 
  NS & NS &  1.4 &    10 &    1.00 &  6.68 & 13.37 \\
  BH & BH &  4.5 &    13 &    4.18 &  6.80 & 13.60 \\
  WD & WD &  0.5 & 11000 &  392.87 &  2.99 &  5.99 \\
  BH & NS &    - &     - &    4.85 &  7.18 & 16.51 \\
  NS & WD &    - &     - &  747.11 &  7.12 &  7.12 \\
  BH & WD &    - &     - & 1966.61 & 10.03 & 10.04
  \enddata
\end{deluxetable}

In what follows, we focus on the collision and tidal capture rate
between two neutron stars. Since we assume a homogeneous distribution
of stars in the core, these results can be re-scaled for close
encounters between other types of stellar objects. In
Table~\ref{tab:rates}, we give these scaling factors in case of a
neutron star (NS), a stellar black hole (BH), and a white dwarf (WD),
with (typical) masses and radii as indicated.  After taking into
account the relative differences in fractional numbers $f_i$, the
collision rate follows from multiplying the (default) NS-NS collision
rate by the factor $\xicol$ in column 5. In turn, multiplying the
resulting collision rate with the factor in column 6 yields an
estimate of the tidal capture rate. Here we neglect the weak
dependence of $\Rmin$ in equation~\eqref{eq:Rmintid} on $\sigc$, which
in turn varies only mildly during the evolution of the core.
While in general a tidal capture does not necessarily lead to the
coalescence of the two stellar objects \citep{lee86}, in case of
compact objects the approach is so close (see $\Rmin/R_2$ in column 7)
that they will merge well within a Hubble time. In particular, for two
neutron stars the close encounter rate is boosted by a factor $\simeq
6.7$ due to tidal capture, without any significant delay with respect
to collisions because of the very efficient emission of gravitational
waves.

The encounter rate might be significantly enhanced due to interactions
between single stars and stellar binaries (and even between binaries),
which have a much larger cross section, with $\Rmin$ proportional to
the binary separation. Moreover, due to mass segregation the more
massive stellar objects, including binaries, sink towards the center
and can be important or even dominant in the core. Still, only the
stars in very ``hard'' binaries with small enough separations (and
thus smaller cross sections) might coalesce within a Hubble time. We
discuss binaries in Section~\ref{sec:compactbinaries} below, in particular
focussing on compact binaries as they are commonly believed to be the
progenitors of SGRBs. We restrict however the encounter rate
calculations to collisions and tidal captures between single stars,
and hence consider the resulting values as lower limits.

\subsection{Evolving encouter rate}
\label{sec:evolution}

After gradual contraction during an early phase, the core of a GC can
go into deep self-similar collapse \citep{lbe80}. The collapse is
halted due to engery release from interactions with binaries, also
known as binary burning \citep{hills75}, and/or from other sources
including a possible intermediate-mass black hole \citep{shapiro77}.
The post-collapse evolution has been extensively studied \citep[see
  e.g.][]{heggie89}, but many aspects are still intensively
investigated and debated.  For example, \cite{fregeau2008} recently
proposed that the observed post-core-collapse GCs ($\simeq 20$\,\% in
the Milky Way) are most likely in the binary burning phase, while the
remaining GCs are still in the process of core contraction and have
not yet reached the binary burning phase. In the meantime, matching
Monte Carlo models of M\,4 \citep{heggiegiersz2008} and of NGC\,6397
\citep{gierszheggie2009} seem to reveal that both GCs are past
collapse in the binary burning phase, even though only NGC\,6397 is
one of the observed post-core-collapse GCs, while the surface
brightness of M\,4 is well fitted by a standard King
profile. \cite{gierszheggie2009} argue that the differences in the
surface brightness profiles are most likely due to fluctuations in the
core after collapse. They point out that it is well possible that most
GCs are post-core-collapse, but only a fraction happens to be at the
``peak" of the fluctuations that corresponds to a cusped surface
brightness profile and leads to post-core-collapse classification.
These fluctuations are not necessarily the well-known gravothermal
oscillations \citep{bettwieser84,goodman87}, but could be the result
of the stochastic nature of binary burning.

Even though it is clear that the details of these fluctuations are far
from known \citep[but see][]{heggiegiersz2009}, the core after
collapse is expected to be \emph{on average} larger than those of the
observed post-core-collapse GCs, and to increase over time to explain
the overall dimming in the central surface brightness \citep[see also
Fig.~10 of][]{gierszheggie2009}.
To mimick this average behaviour, we assume a gradually expanding core
after deepest core collapse, which in turn we describe by two
classifcal self-similar solutions in two phases.
In the early phase expansion mirrors the self-similar late phase
collapse, but at a slower rate \citep{ilb83}.  After a time since
deepest collapse roughly equal to the duration of the late phase
collapse \citep{heggie85}, the core enters the late phase in which it
follows the self-similar expansion of an isolated system
\citep{henon65}. We derive the encounter rate during all four phases
of collapse and expansion.

In addition to this core evolution model with gradual core expansion
after deepest collapse, we also consider a model in which the core
properties are kept fixed after deepest collapse.
Since the latter ``halt of collapse" would mimick binary burning
without fluctuations, it provides an upper limit to the contribution
of high-density cores, and hence an upper limit to the encounter
rate. 
On the other hand, while the core expansion is just an average
approximation to the possible complex fluctuations after deepest
collapse, we believe it to provide a closer estimate to the encounter
rate. 
Moreover, we find below (see also Figure~\ref{fig:m15_trh_frac}) that
the predicted fraction of (post-)core-collapse GCs is similar to that
observed for the MW. The core expansion could also naturally explain
why only loose GCs have a shallow/depleted global mass function
\citep{demarchi2007}.

We adopt M\,15 as a proto-typical GC that underwent core collapse.
Because this GC has been extensively studied and modelled
\citep{dull97, mcnamara04, bosch06}, its current properties are known
accurately. Given the high concentration $c = \log(\rc/\rt) \gtrsim
3$, where $\rt$ is the tidal radius, we assume that the core of M\,15
is currently still very close to deepest collapse. The fraction of its
age $\tage \simeq 13.2$\,Gyr \citep{baumgardt03} that M\,15 spent in
the early and late phase of core collapse depends on its initial
concentration $\cini$. The latter can be inferred from evolved
(single-mass) Fokker-Planck models \citep{quinlan96} for a given
initial half-mass relaxation time $\trhini$. In turn, the latter
follows from the current half-mass relaxation time $\trh \simeq
1.42$\,Gyr as $\trhini \simeq \trh(\tage) - \tage/\xides$, under the
assumption of a constant mass loss rate from the GC
\citep[e.g.][]{vesperini97}. Since the rate of dissolution, $\xides$,
depends again on the initial concentration \citep{gnedin99}, we have
to find $\cini$ iteratively.

Doing this, we infer for M\,15 $\xides \simeq 51$, $\trhini \simeq
1.68$\,Gyr, and corresponding $\cini \simeq 1.70$. This implies that
M\,15 spent about $8.0$\,Gyr in the early phase, and took another
$5.2$\,Gyr to reach the current phase of deep core collapse. Even
though the encounter rate during the late phase is expected to
dominate over the encounter rate during the early phase, we still take
the increase in $\nuenc$ during the early phase into account. In
particular, we take into account the changes in the fractions of
stellar types due to mass segregation in the early phase, whereas we
assume the fractions to remain constant during deep collapse. We
predict the core expansion by mirroring the late phase collapse but
with an expansion rate that is a factor 3 slower than the collapse
rate. After about $5.2$\,Gyr (the duration of the late phase collapse)
in the future, we expect the core of M\,15 to go into the late
self-similar expansion phase.

\begin{figure}
\includegraphics[width=\columnwidth,angle=0,scale=1.]{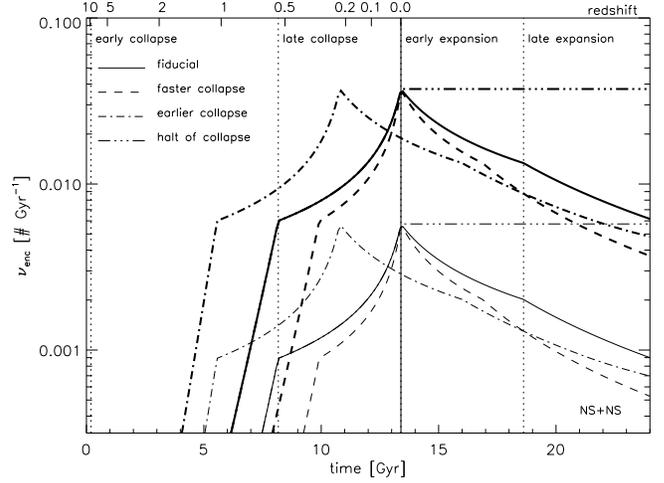}
\caption{An estimate of the encounter rate between two neutron stars
  in M\,15 as function of time (or redshift at the top). The thin
  lower curves show the predictions for the collision rates, while the
  thick upper curves present the tidal capture rates.  The solid
  curves are for the fiducial core evolution model, whereas the dashed
  curves assumes a faster collapse (and expansion), and the
  dash-dotted curves indicate the effect of an earlier collapse. The
  dash-triple-dotted curve shows an additional model in which the core
  properties are kept fixed after deepest collapse. (See
  Sections~\ref{sec:evolution} and~\ref{sec:uncertainties} for further
  details.)}
\label{fig:m15_nuencevo}
\end{figure}

Figure~\ref{fig:m15_nuencevo} presents the resulting rates as function
of time (and redshift along the top axis) for close encounters between
two neutron stars in the core of M\,15. The number fraction of neutron
stars in the core at the beginning of the early phase is less than
$1$\%, but it rises due to mass segregation to about $55$\% during the
self-similar deep collapse \citep{dull97}. The thin lower curves show
the predictions for the collision rates, while the upper thick curves
present the tidal capture rates. The solid curves are for the fiducial
core evolution model, whereas the dashed curves assumes a faster
collapse (and expansion), and the dash-dotted curves indicate the
effect of an earlier collapse. These two variations on the fiducial
model are further discussed in Section~\ref{sec:uncertainties}
below. Finally, the dash-triple-dotted curve shows the additional
model in which the core properties are kept fixed after deepest
collapse.

\subsection{Average encounter rate}
\label{sec:average}

We compute an average encounter rate per host galaxy by assuming that
all GCs follow an evolution similar to that of M\,15, with a scaling
in time, based on the half-mass relaxation time $\trh \propto
\rh^{3/2} M^{1/2}$. We adopt the distribution in half-mass radii $\rh$
of GCs derived by \citet{jordan05}, which is independent from the
distribution in their total masses $M$ \citep{mclaughlin00}. It is
believed that initially the GC mass distribution followed a power-law,
but that especially the less-massive GCs dissolved, most, if not all,
before going into core collapse \citep{mclaughlin08}. Hence, we adopt
the current GC mass distribution, which is well described by a
lognormal distribution \citep{jordan06}.
After randomnly drawing from these distributions in $\rh$ and $M$, we
arrive at a current half-mass relaxation time, i.e., after a time
$\tage \simeq 13$\,Gyr since the formation of an average old GC.
Adopting the same initial concentration $\cini = 1.70$ as estimated
for M\,15, we infer the corresponding initial half-mass relaxation
time, which we use to re-scale the above derived encounter rate
evolution for M\,15.

\begin{figure*}
  \includegraphics[width=\columnwidth,angle=0,scale=1.0]{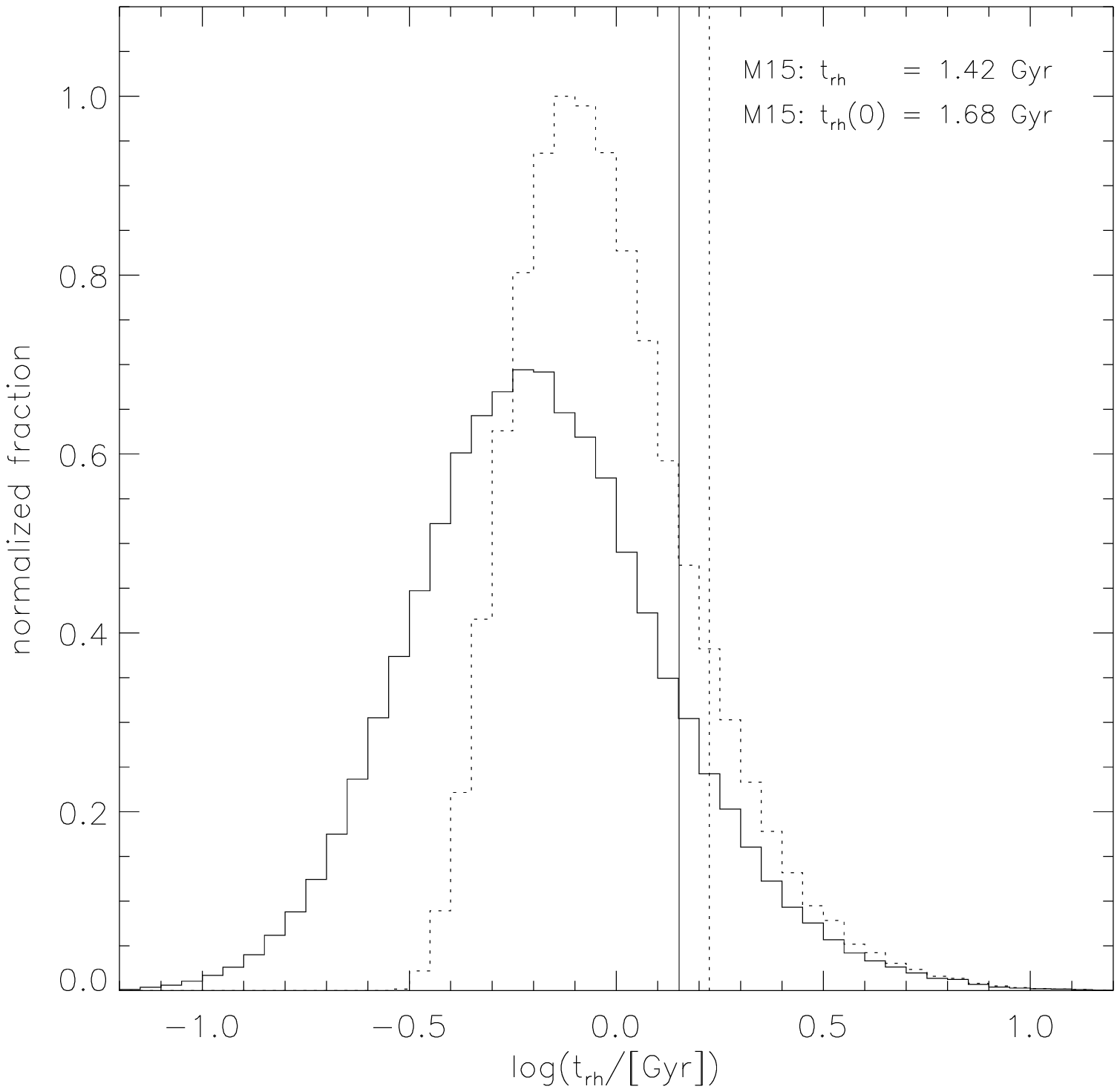}
  \hfill
  \includegraphics[width=\columnwidth,angle=0,scale=1.0]{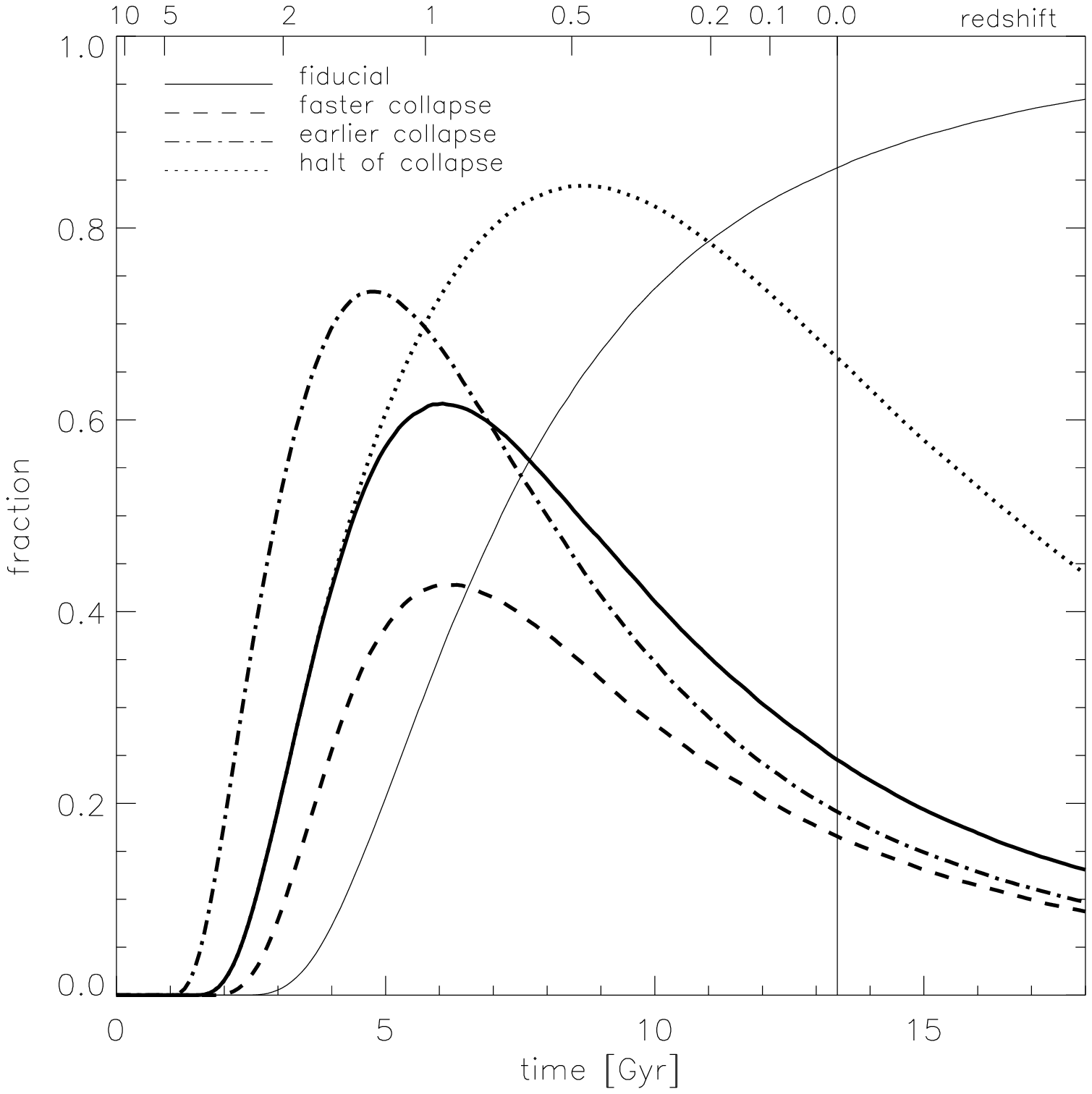}
  \caption{The \emph{left panel} shows the distribution of half-mass
    relaxation times $\trh \propto \rh^{3/2} M^{1/2}$, based on the
    independent distributions in half-mass radii $\rh$ and total mass
    $M$ of globular clusters. This results in the solid histogram,
    which turns into the dotted histogram for initial half-mass
    relaxation times $\trh(0)$. The vertical lines show that M\,15 has
    a larger than average half-mass relaxation time, so we expect that
    currently more than half of all GCs are already past reaching
    deepest core collapse. This is confirmed by the thin solid curve
    in the \emph{right panel}, showing the fraction of GCs that past
    reaching deepest core collapse as function of time. However, in
    this fiducial core evolution model it is likely that some time
    after deepest core collapse the GCs are not anymore
    \emph{observed} to be post-core-collapse because the core is
    expanding. Henceforth, to estimate the fraction of observed
    post-core-collapse GCs, we only count at a given time GCs that are
    in the late collapse or early expansion phase when the core is
    smaller. The resulting thick curves show that for the first three
    models the current fraction (vertical solid line) is similar to
    the fraction of $\simeq 20$\,\% of post-core-collapse GCs observed
    in the Milky Way. Only the fourth halt-of-collapse model predicts
    a much higher fraction because the core properties are held fixed
    after deepest collapse.}
\label{fig:m15_trh_frac}
\end{figure*}

The current and initial half-mass relaxation distributions are shown
as solid and dotted histograms in the left panel of
Figure~\ref{fig:m15_trh_frac}, while the vertical lines indicate the
corresponding values for M\,15.
Since M\,15 has a larger than average half-mass relaxation time, we
expect that currently more than half of all GCs are already past
reaching deepest core collapse.  This is confirmed by the thin solid
curve in the right panel of Figure~\ref{fig:m15_trh_frac}, showing the
fraction of GCs that past reaching deepest core collapse as function
of time. 
In this fiducial core evolution model it is likely that some time
after deepest core collapse the GCs are not anymore \emph{observed} to
be post-core-collapse because the core is expanding. Henceforth, to
estimate the fraction of observed post-core-collapse GCs, we only
count at a given time GCs that are in the late collapse or early
expansion phase when the core is smaller. The resulting thick curves
show that for the first three models the current fraction (vertical
solid line) is similar to the fraction of $\simeq 20$\,\% of
post-core-collapse GCs observed in the Milky Way. Only the
halt-of-collapse model predicts a much higher fraction because the
core properties are held fixed after deepest collapse.

\begin{figure}
\includegraphics[width=\columnwidth,angle=0,scale=1.]{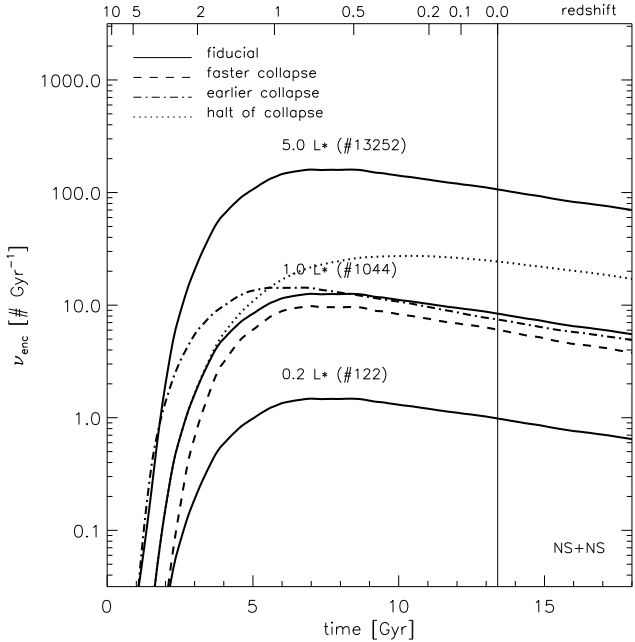}
\caption{An {\em average} encounter rate between two neutron stars in the
  core of globular clusters per host galaxy with luminosity $\Lg =
  \{0.2,1.0,5.0\} \times \Ls$. The expected number of GCs per host
  galaxy, given in brackets, follows from \citet{mclaughlin99}. For
  each globular cluster, the collision rate of M\,15 is re-scaled by
  the half-mass relaxation time $\trh \propto \rh^{3/2} M^{1/2}$,
  randomly drawn from the independent distributions in half-mass radii
  $\rh$ and total mass $M$ of globular clusters. The meaning of
  different line styles is the same as in Fig.\ref{fig:m15_nuencevo}.}
\label{fig:m15_nuencscaled}
\end{figure}

The number of GCs for which we re-scale the encounter rate evolution
of M\,15 depends strongly on the total luminosity $\Lg$ of the host
galaxy \citep{mclaughlin99}.
In Figure~\ref{fig:m15_nuencscaled} we plot the combined encounter
rate for a typical $\Ls$ galaxy with about a thousand GCs, as well as
for a galaxy that is a factor $5$ less and more luminous with the
number of GCs comparable to that in the Milky Way and in a giant
elliptical galaxy, respectively. In the latter two cases we show only
the tidal capture rates for the fiducial model, wheras for the $\Ls$
galaxy we also indicate the effect of a faster and earlier collapse as
discussed in Section~\ref{sec:uncertainties}, as well as the
additional half-of-collapse model. The results for the collision rate
are not shown for clarity, but are similar except for a factor $\simeq
6.7$ decrease in amplitude. In all cases, the encounter rate increases
strongly up to redshift $z \sim 1$, but slowly decreases again toward
$z = 0$.

Next, we estimate a global encounter rate by combining the previously
derived encounter rates per host galaxy with the galaxy luminosity
density distribution. The latter is well described by the Schechter
function 
\begin{equation}
  \label{eq:gallumdistr}
  \Phig(L) = (\Phigs/\Ls) \; (L/\Ls)^\alpha \exp(-L/\Ls),
\end{equation}
with $\Phigs = 3.77 \times 10^{-3}$\,\Mpcc\ and $\alpha = -1.30$
\citep{faber07}. Since the above estimate of the combined encounter
rate evolution depends on the {\em current} luminosity of the host
galaxy, the evolution of $\Phig(L)$ is not needed to compute the
global encounter rate as function of time. However, the latter
evolution should be taken into account when discussing the properties
of host galaxies, which do change significantly with redshift (see
Section~\ref{sec:galacticenvironment}).

\begin{figure}
\includegraphics[width=\columnwidth,angle=0,scale=1.]{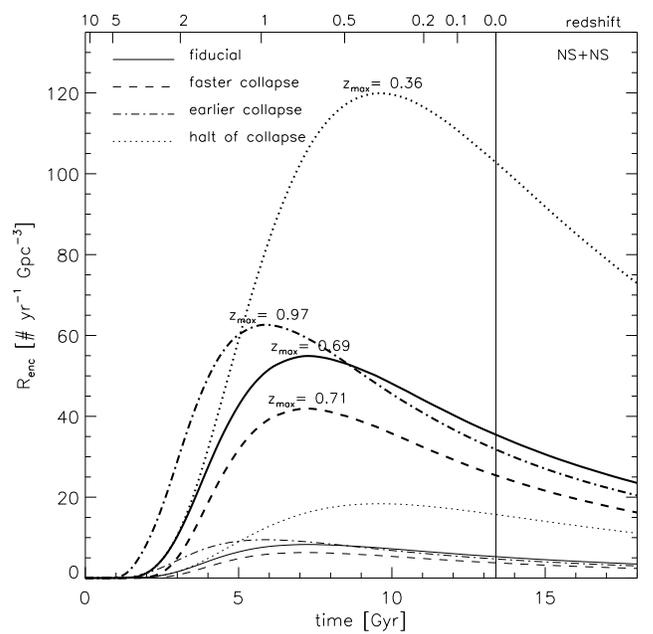}
\caption{The predicted {\em global} rate of close encounters between two
  neutron stars in the cores of globular clusters, per volume of space and as
  function of time (or redshift at the top). The values are computed
  from combining the average collision rate per host galaxy
  (Fig.~\ref{fig:m15_nuencscaled}) with the galaxy luminosity density
  distribution. The meaning of different line styles is the same as in
  Fig.\ref{fig:m15_nuencevo}.}
\label{fig:m15_nuencspace}
\end{figure}

In Figure~\ref{fig:m15_nuencspace}, we present our prediction for the
global encounter rate $\Renc$ (in \pyrGpcc) between two neutron stars
in the core of GCs. The rates both of collisions (thin curves) and of
tidal captures (thick curves) show a clear increase around $z \sim 1$,
followed by a gradual decline. The tidal capture rate for the fiducial
model (thick solid curve) peaks around $z \simeq 0.7$, at a value of
$\Renc \simeq 55$\,\pyrGpcc. This value is fully consistent with the
estimated event rates for SRGBs, which are of the order
8--30\,\pyrGpcc\ for isotropic radiation \citep{guetta06}. Collimated
radiation with opening angles of 30--60\dgr, would boost up the event
rates of SGRBs to our predicted global encounter rates. However, since
the latter are likely to be rather conservative, smaller opening
angles would still be possible.

\subsection{Faster and ealier collapse}
\label{sec:uncertainties}

To infer the initial concentration of M15 and the corresponding time
spent in the early and late phase of collapse, we used the results of
a series of cluster models with single-mass stars \citep{quinlan96}.
Several complicating factors in real clusters can change the rate of
collapse \citep[see e.g.\ review by][]{chernoff93}: for example
stellar evolution and especially the presence of primordial binaries
slow down the collapse, while mass loss from a tidal boundary and in
particular mass segregation accelerate the collapse.
Even though the fraction of primordial binaries in GCs is believed to
be only a few per cent \citep[e.g.][]{hurley07, davis08}, the binding energy of
a single hard binary is sufficient to significantly slow down or
perhaps even prevent core collapse \citep{hut92, heggie06}.
On the other hand, because the relaxation time is inversely proportional
to the mean stellar mass, the acceleration of core collapse due to
mass segragation is roughly proportional to the increase in the mean
mass, which in the core of M\,15 is about a factor $1.19/0.43 \sim 3$
\citep{dull97}.

The detailed inclusion of these competing mechanism is beyond the
scope of this paper, but to investigate the effect of a changing
collapse rate on the encounter rate, we assume a 50\% faster core
collapse (and expansion) than the fiducial M\,15 core evolution model.
At the same time, this implies a lower initial concentration of $\cini
\simeq 1.47$. 
The results of this ``faster collapse'' model on the encounter rate
are shown with dashed curves in
Figures~\ref{fig:m15_nuencevo}--\ref{fig:m15_nuencspace},
and~\ref{fig:m15_nuenccumul}. As can be seen from
Figure~\ref{fig:m15_nuencspace}, the main effect is a lowering of the
global encounter rate by about 24\% with respect to the fiducial model
to $\Renc \simeq 41$\,\pyrGpcc\ at the peak, which remains around $z
\sim 0.7$.

We already mentioned in Section~\ref{sec:evolution} that after deepest
core collapse the core can undergo fluctuations, either due to
gravothermal oscillations \citep{bettwieser84,goodman87}, or as a
result of the stochastic nature of binary burning
\citep{gierszheggie2009}. In either case, these fluctuations can
result in change in the core radius $\rc$ by factors of a few, and
even order of magnitude changes in the core density $\rhoc$.
We do not attempt to capture this fluctuating nature, but we consider
here a possible consequence of it. Because the (central) relaxation
time is inversely proportional to $\rhoc$, only a relatively small
fraction of the time during an oscillation will be spent at large
$\rhoc$ and correspondingly small $\rc$. Still, if we happen to observe
a post-collapse GC during this special time at the ``peak" of the
fluctuation, we will underestimate the time elapsed since deepest
collapse under our assumption of a steady expansion in our fiducial
core evolution model.
To investigate the possible effect on the encounter rate, we consider
a shift the time of deep core collapse back in time by an amount of
$2.6$\,Gyr, i.e., half of the duration of the late-phase collapse. As
a consequence, the early-phase collapse is shorter by the same amount
and the initial concentration increases to $\cini \simeq 1.79$. The
smooth expansion model in this case predicts a current core radius and
core density that are respectively almost 65\% larger and a factor 3
smaller than the observed values of $\rc = 0.05$\,pc \citep{noyola06}
and $\rhoc = 6.5 \times 10^6$\,\Msunpcc\ \citep{bosch06}.
The results of this ``earlier collapse'' model on the encounter rate
are indicated by the dash-dotted curves in
Figures~\ref{fig:m15_nuencevo}--\ref{fig:m15_nuencspace},
and~\ref{fig:m15_nuenccumul}. Figure~\ref{fig:m15_nuencspace} shows
that the global encounter rate is increased by about 14\% with respect
to the fiducial model to $\Renc \simeq 62$\,\pyrGpcc\ at the peak,
which occurs earlier, at $z \sim 1.0$.
The latter is directly related to the applied shift in time, whereas
the former is the result of more GCs reaching deep collapse in $\tage
= 13$\,Gyr (see also Figure~\ref{fig:m15_trh_frac}).

The above faster and earlier collapse models imply respectively a
lower and higher initial concentration for M\,15 than used in the
fiducial model. For each model, we then fixed the initial
concentration of all GCs to that of M\,15 when computing the average
encounter rate. It is already very difficult to infer the initial
concentration for M\,15, let alone the distribution of initial
concentrations for the population of old GCs. Even so, we have
recomputed the average encounter rate adopting a Gaussian distribution
in $\cini$, with a mean at the initial concentration of M\,15 and with
dispersion in the range $[0,0.3]$. As a result of the dispersion, the
rate of dissolution $\xides$ is not anymore fixed to that of M\,15,
which in turn leads to an additional broadening in the half-mass
relaxation distribution when converting from current $\trh$ to initial
$\trh(0)$ values. However, since the resulting variations in the
conversion are typically less than $\sim 0.1$\,Gyr, the effect on the
average encounter rate is very small. Changing the mean, as we do in
the faster and earlier collapse models, does have a significant effect
on the average encounter rate. Nevertheless, we see from
Figure~\ref{fig:m15_nuencspace} that the evolution of the global
encounter rate is quite similar even for significant deviations from
the fiducial mdoel. In other words, as long as M\,15 is indeed
representative of an (old) GC that underwent core collapse and
expansion, we expect our predictions for the encounter rates to be
robust.

\subsection{Binaries}
\label{sec:compactbinaries}


We have not included the potentially important effect of binaries in
our encounter rate calculations. The fraction of primordial binaries
in GCs is thought to be only a few per cent \citep{hurley07}, as
confirmed by measuring photometric binaries beyond the half-light
radius \citep[see e.g.][]{davis08}. In the core of GCs tidal captures
can significantly increase the fraction of binaries. Placing useful
observational constraints on the fraction of binaries in the core is
not an easy task because many of them could be dark as a result of
binary-single interactions in which the more numerous main sequence 
stars (MS) and white dwarfs (WD) are replaced by compact
remnants. Binary systems have a larger cross section for intersection
and they mass segregate faster into the denser core than individual
stellar objects. Three-body interactions are thus indeed likely to
create compact binaries in the cores of GCs, including double neutron
star (NS-NS) and black hole--neutron star (BH-NS) systems. As already noted in the Introduction, 
the merger of such dynamically formed compact binaries is potentially an 
important channel for the production of SGRB progenitor systems
\citep{grindlay06}, different from the close encounters between single
compact stellar objects we have considered so far. As we discuss below in \ref{sec:dynamicalcompactbinaries}, 
for encounters involving the exchange of a main sequence star with a neutron star or a black hole, the resulting separation will be too large for coalescence to take place in less than a Hubble time. Thus this channel is unlikely to contribute to the overall formation rate of viable SGRB progenitors. The occurrence of a direct impact may release a substantial amount of energy, but whether this can lead to the conditions required for SGRB production remains to be evaluated.

The three main ways to arrive at compact binaries are (i) primordial
compact binaries, (ii) primoridal binaries of which the non-compact
member(s) are replaced by compact objects via one (or two) exchange
interactions, and (iii) the same as (ii) but starting from binaries
which themselves are first formed through tidal capture of two single
stellar objects.
We consider all three ways below and show that in the cores of GCs,
the latter dominates the formation rate of compact binaries, but is
not, as commonly believed, dominant over the rate of close encounters
between two single compact objects.
While SGRBs are potentially created instanteneously when two compact
objects closely encounter each other, the separation of the compact
binary has to be small enough such that the time for the compact
objects to merge, in addition to the formation time, is less than the
Hubble time.

Note that we expect very little contribution to the close encounter
rate from interactions between a single compact object (NS or BH) and
a compact object that is a member of a binary which also contains a
extended object (MS or WD). Even though we show below that the
fraction of the latter binaries can be(come) significant, in nearly
all cases the (exchange) interaction will be with the extended object
\citep[e.g.][]{sigurdsson93}.

\subsubsection{Primordial compact binaries}
\label{sec:primordialcompactbinaries}

The formation of primordial binaries follows the cosmic star formation
rate which peaks at high redshift $z\sim3$ \citep[e.g.][]{hopkins06}.
Primordial binaries with two massive stars ($\gtrsim 8$\,\Msun) that
survive the mass loss and possible asymmetry of the supernova
explosion of the secondary provide compact NS-NS and BH-NS binaries.
They potentially could produce SGRBs at high redshift with a
distribution extending to lower redshift depending on the range of
initial binary separations and thus merger times.  
However, even if a non-neglible fraction of primordial massive
binaries survive and result in close compact binaries \citep[$\sim
10^{-2.5}$,][]{narayan91}, massive binaries are only a very small
fraction of all primordial binaries for a reasonable, not too
top-heavy initial mass function ($\sim 10^{-4.5}$, for an IMF with a
Salpeter power law slope from $100$\,\Msun\ down to $0.3$\,\Msun, and
constant down to $0.1$\,\Msun).  In GCs with already at most a few per
cent primordial binaries \citep[$\sim 10^{-1.5}$,][]{hurley07,
  davis08} and likely additional complications to form and retain
compact binaries \citep[see also][]{ivanova08}, their possible
contribution to producing SGRBs should be negligible; the GCs in the
Milky Way contain a total of $\sim 10^{7.5}$ stars, which implies that
we expect only $\sim 1$ primordial compact binary formed in the GCs of
the Milky Way.

However, in the ``field'' of a galaxy the primordial binary fraction
is of order unity and the vast amount of them results in a number of
compact binaries that is not anymore insignificant; we expect $\sim
10^5$ primordial compact binaries formed in the ``field" of the Milky
Way \citep[see, e.g.][for initial estimates]{narayan91,phinney91}. Merger rate calculations for such primordial binaries in the galactic field have become increasingly sophisticated \citep{kalogera04,osh05,osh08,osh10}, and now consider not only various stellar evolutionary channels, but also different host galaxy types (spiral vs. elliptical). The results are still hindered by the fact that there is a small number of observed systems, giving broad estimates in the range 1--100\,\pyrGpcc. 

\subsubsection{Dynamically formed compact binaries}
\label{sec:dynamicalcompactbinaries}

Compact binaries formed dynamically in GCs have long delay
times between compact object birth and binary formation as they have
to wait for collapse of the core \citep{hopman06}. This helps their formation in three
ways: (i) mass segregation increases the relative fraction of
heavier-than-average binaries as well as compact remnants in the core,
(ii) tidal capture increases the binary fraction in the core, in
particular the ``hard'' binaries with small separations (whereas the
``soft'' binaries with large separations are being ``ionized''), (iii)
the high core density allows three-body exchange interactions in which
a member of an existing binary is replaced with a compact remnant.
Henceforth, we expect in the cores of GCs the compact binary formation
rate to follow the close encounter rate, with a similar peak around $z
\simeq 0.7$ as shown in Figure~\ref{fig:m15_nuencspace}. Whereas after
this initial delay the potential production of SGRBs is nearly
instantaneous in the proposed encounter scenario, there is an
additional delay in the binary merger scenario given by the rate of
angular momentum losses through gravitational waves. 

For a NS-NS binary dynamically formed around $z\simeq 0.7$ to merge
within the next $\simeq 6$~Gyr through the emission of gravitational
waves, the binary separation required is $a \lesssim
3.93$\,\Rsun.
As mentioned above, most of the compact binaries in GCs are expected
to be formed through three-body exchange interactions: a NS (average
mass $1.4$\,\Msun, and average radius $10$\,km) replaces the
less-massive main-sequence star (MS; $0.4$\,\Msun, $3.3 \times
10^5$\,km) or white dwarf (WD; $0.5$\,\Msun, $1.1 \times 10^4$\,km)
member in an existing binary with a NS. 
The latter NS-MS or NS-WD binary might be itself the result of a
previous similar three-body exchange interaction with a MS-MS, WD-MS
or WD-WD binary.
We assume as minimum separation of these binaries the closest distance
before ignition of mass transfer occurs \citep{paczynski71}. This
yields $a \gtrsim 1.72$\,\Rsun\ and $a \gtrsim 0.054$\,\Rsun\ for
NS-MS and NS-WD binaries, respectively. We obtain similar lower limits
of about $1.312$, $1.365$, and $0.043$ in units of \Rsun\ for MS-MS,
WD-MS, and WD-WD binaries, respectively.

Even though after each exchange the binary hardens, the binary
separation after an exchange, $a_\mathrm{fin}$, is typically larger
than the binary separation before the exchange, $a_\mathrm{ini}$.
\cite{sigurdsson93} show that the distribution of
$a_\mathrm{fin}/a_\mathrm{ini}$ strongly peaks around the mass ratio
$m_f/m_e$, between the single stellar object from the ``field'' and
the stellar object ``exchanged'' from the binary, but with a long tail
towards lower values. 
Taking for $a_\mathrm{ini}$ the minimum separation of each binary, we
compute the median of the latter distribution\footnote{We adopt an
  approaching velocity from infinity of $v_\infty = 2\sigc \simeq
  20$\,\kms, but the results are not sensitive to the precise value
  that is adopted.}  to arrive at the average $a_\mathrm{fin}$ after
the first exchange interaction, and repeat this for the MS-MS, WD-MS,
and WD-WD binaries that require a second exchange interaction to
arrive at a NS-NS binary.
For all three binaries with a MS as a member the resulting
\emph{minimum} binary separations ($5.21$, $12.8$ and $12.4$ in units
of \Rsun\ for NS-MS, MS-MS, and WD-MS, respectively) are larger than
the above \emph{maximum} binary separation for merger of the
dynamically formed NS-NS binary ($3.93$\,\Rsun\ for merging within
$\simeq 6$\,Gyr). Note that this is still the case even when we ignore
the time to reach core collapse and allow the merger to take a full
Hubble time, corresponding to a maximum binary separation of
$4.8$\,\Rsun.
On the other hand, a NS-WD binary results in a NS-NS binary with a
minimum separation $a \gtrsim 0.126$\,\Rsun\ well below the maximum
for merging. Similarly, a WD-WD binary yields a NS-WD binary with $a
\gtrsim 0.102$\,\Rsun\ after the first exchange, and a NS-NS binary
with $a \gtrsim 0.240$\,\Rsun\ after the second exchange.

We assume that the formation rate of these binaries through three-body
exchange interactions follows that of the collision rate, but with a
different amplitude, which we estimate below.

\subsubsection{Compact binaries from primordial binaries}
\label{sec:fromprimordialbinaries}

Here we estimate the rates at which compact binaries in the cores of
GCs are dynamically formed from primordial binaries, of which one or
both members are non-compact stellar objects.
We take $f_\mathrm{pri} = 0.02$ for the total fraction of primordial
binaries in GCs, which is not well known but thought to be at most a
few per cent \citep[e.g.][]{hurley07, davis08}.
To derive the fraction of primordial binaries that have a NS and a WD,
we adopt an IMF with Salpeter power-law slope from $100$\,\Msun\ down
to $0.3$\,\Msun, and constant down to $0.1$\,\Msun. Supposing that
stars above $20$\,\Msun\ evolve into black holes (BHs), stars above
$7$\,\Msun\ turn into SNe, stars above $0.8$\,\Msun\ end their life as
WDs, and lower-mass stars are still in the MS, we find corresponding number
fractions of about $0.0012$ $0.0043$ $0.1014$, and $0.8930$ for BH,
NS, WD, and MS stars, respectively.
While we start with drawing the masses of the primary and secondary,
$M_1$ and $M_2$, indepdently from the IMF, we also consider the case
that the mass ratio $M_2/M_1$ is distributed uniformly in the range
from zero to unity.
The former case is fully consistent with observed (spectroscopic)
binaries with longer periods ($P>10^3$\,days), but there are
indications of a flatter distribution in mass ratios for the
shorter-period binaries, possibly because stars that form close to
each other become closer in mass by reaching an equilibrium in mass
transfer.
We adopt a lognormal distribution in binary periods $P$ (in days) with
mean $\mu_{\log P} = 4.8$ and standard deviation $\sigma_{\log P}=2.3$
\citep{duquennoy91}.

We first consider the formation of NS-NS binaries from primoridal
NS-WD binaries through a single exchange interaction.
With the masses of the primary and secondary independently drawn from
the IMF, we expect a fraction $f_\mathrm{NS-WD,pri} \simeq 8.81 \times
10^{-4}$ of all primordial binaries to be a NS-WD binary.
Together with the stars and compact remnants that are heavier than the
MS stars, the binary sinks to the core of a GC. During deep core
collapse the NS-WD undergoes a three-body exchange interaction with a
NS for which we adopt the number fraction $f_\mathrm{NS} \simeq 0.547$ of
M\,15 \citep{dull97}.
We only count the NS-WD binaries with separations in the range
$[0.054,1.669]$\,\Rsun. The lower limit is set to avoid stable mass
transfer, while the upper limit increases after the exchange
interaction to the maximum separation of $3.93$\,\Rsun\ allowed for
merger of the dynamically formed NS-NS binary within $\simeq 6$\,Gyr.
Given the above lognormal distribution in binary periods, the fraction
of NS-WD binaries in this range of separations is $f_\mathrm{sep}
\simeq 1.83 \times 10^{-2}$. The corresponding mean separation is
$\overline{a} \simeq 0.829$\,\Rsun.
We now obtain the rate from eq.~(\ref{eq:nucoll}) with $\Rmin =
\overline{a} + r_\mathrm{NS} \simeq 0.829$\,\Rsun, $f_1 =
f_\mathrm{pri} \, f_\mathrm{NS-WD,pri} \, f_\mathrm{sep} \simeq 3.23
\times 10^{-7}$, $f_2 = f_\mathrm{NS} \simeq 0.547$, and $m_1 + m_2
\simeq 2 \times 1.4 + 0.5 = 3.3$\,\Msun.
In this way, we predict that the rate $\Rdyn$ to turn a primordial
NS-WD binary dynamically into a NS-NS binary that can merge within the
next $\simeq 6$\,Gyr, is a factor $\Rdyn/\Rcol \simeq 4.02 \times
10^{-2}$ smaller than the collision rate, and a factor $\Rdyn/\Rtid
\simeq 0.618 \times 10^{-2}$ smaller than the tidal capture rate.

We now turn to the additional dynamical formation of a NS-NS binary
from a primoridial WD-WD binary through two subsequent three-body
exchange interactions with a NS. 
We assume that the first exchange to form the NS-WD binary happens
some time before deepest core collapse, with a formation rate that can
be computed in the same way as above. The fraction of primodial
binaries that are WD-WD binaries is about $f_\mathrm{WD-WD,pri} \simeq
1.03 \times 10^{-2}$.  We count the WD-WD binaries with separations in
the range $[0.043,0.708]$\,\Rsun, with the lower limit again to avoid
stable mass transfer, while the upper limit increases after two
subsequent exchange interactions to the maximum separation of
$3.93$\,\Rsun\ allowed for merger of the dynamically formed NS-NS
binary within $\simeq 6$\,Gyr.  This range in separation corresponds
to a fraction $f_\mathrm{sep} \simeq 1.10 \times 10^{-2}$ and mean
separation $\overline{a} \simeq 0.371$\,\Rsun. We then use as before
eq.~(\ref{eq:nucoll}) with $\Rmin \simeq 0.371$\,\Rsun, while we set
$f_1 = f_\mathrm{pri} \, f_\mathrm{NS-WD} \, f_\mathrm{sep} \simeq
2.27 \times 10^{-6}$, and $m_1 + m_2 \simeq 1.4 + 2 \times 0.5 =
2.4$\,\Msun.
As in Section~\ref{sec:evolution}, we assume $f_2 = f_\mathrm{NS}(t)$ to
vary from its primordial value to $0.547$ during the early-phase
collapse, and then to remain the same during the late-phase collapse.
We intergrate the resulting rate of exchange interactions between a
primordial WD-WD binary and single NS over the full duration of the
core collapse phase. This yields the expected number ($\simeq 4.24
\times 10^{-4}$) of dynamically formed NS-WD binaries with separations
within the above range.
We then divide this number by the number of stellar systems ($\simeq
2.86 \times 10^3$) in the core $N_c = (4\pi/3) n_c r_c^3$ at deepest
core collapse to arrive at a fraction $f_\mathrm{NS-WD,dyn} \simeq
1.48 \times 10^{-7}$.
Next, we repeat the rate computation but now for dynamically forming a
NS-NS binary in a second exchange interaction with $f_1 =
f_\mathrm{NS-WD,dyn}$, $f_2 = f_\mathrm{NS} \simeq 0.547$, $m_1 + m_2
\simeq 2.4$\,\Msun, and $\Rmin = 0.873$\,\Rsun. The latter value
arises from the above mean separation $\overline{a} \simeq
0.371$\,\Rsun\ after the first exchange interaction, multiplied by the
(median) increase after the second exchange interaction. 
In this way, we find that the rate to dynamically form a NS-NS binary
from a primordial WD-WD binary through two subsequent three-body
exchange interactions is smaller than the collision rate by a factor
$\Rdyn/\Rcol \simeq 1.94 \times 10^{-2}$, and smaller than the tidal
capture rate by a factor $\Rdyn/\Rtid \simeq 0.299 \times 10^{-2}$.

By combining the above two formation channels, we thus expect that the
rate to dynamically turn primordial binaries into compact NS-NS
binaries that can merge in time is smaller than the tidal capture rate
by a factor $\simeq 0.916 \times 10^{-2}$.
In case that the mass ratio $M_2/M_1$ of the primordial binary is
uniformly distributed, the fraction of primordial NS-WD and WD-WD
binaries increases to $f_\mathrm{NS-WD,pri} \simeq 4.76 \times
10^{-3}$ and $f_\mathrm{WD-WD,pri} \simeq 7.18 \times 10^{-2}$, i.e.,
a factor $\simeq 5.41$ and $\simeq 6.98$ higher than when $M_1$ and
$M_2$ are independently drawn. As a result, the rate to form compact NS-NS
binaries increases by a factor $\simeq 5.92$, but is still only a
fraction $\simeq 5.43 \times 10^{-2}$ of the tidal capture rate.

A similar calculation for compact BH-NS binaries dynamically formed
from primordial binaries shows that also in this case the minimum
separations of primordial binaries with a MS star as a member are
larger than the maximum separation of $\lesssim 6.34$\,\Rsun\ for
which the BH-NS can merge within $\simeq 6$\,Gyr. However, BH-NS
binaries can form dynamically from primordial NS-WD or BH-WD binaries
via a single exchange interaction with respectively a BH or NS, or
from primoridal WD-WD binaries via two subsequent exchanges
interaction with both a NS and a BH. We compute the rate for each of
these formation channels in the same way as above. We assume that the
fraction of BHs relative to the (changing) fraction of NSs remains
equal to the primoridal fraction, i.e., $f_\mathrm{BH}/f_\mathrm{NS}
\simeq 0.283$ throughout. 
If instead of chosing the masses of the primary $M_1$ and secondary
$M_2$ of the primordial binary indepedently, we assume that the mass
ratio $M_2/M_1$ is uniformly distributed, the fraction of primordial
NS-WD, BH-WD and WD-WD binaries increases by a factor $5.41$, $1.73$,
and $6.98$, respectively.
At the end, the combination of the three formation channels yields a
formation rate of compact BH-NS binaries is smaller than the tidal
capture rate between a single BH and NS by a factor $\simeq 0.593
\times 10^{-2}$ if $M_1$ and $M_2$ are independently drawn, and a factor
$\simeq 1.88 \times 10^{-2}$ when $M_2/M_1$ is uniformly distributed.

\subsubsection{Compact binaries from tidally captured binaries}
\label{sec:fromtidallycapturedbinaries}

It is clear from the above calculations that in the cores of GCs close
encounters between single compact objects are significantly more
likely than mergers of compact binaries that are dynamically formed
from primordial binaries. When starting instead from binaries that are
formed through tidal captures the chance of dynamically forming and
merging a compact binary can be much higher as we show below.
As before binaries with a MS as a member are omitted, because they
result after exchange interaction(s) in compact binaries with minimum
separations that are larger than the maximum separation allowed for
them to merge in time.

We first consider the formation of NS-NS binaries from tidally
captured NS-WD binaries.
To compute the fraction of tidally captured NS-WD binaries, we
integrate the tidal capture rate between a NS and WD over time until
deepest core collapse, and then divide by the number of stellar
systems in the core.
The tidal capture rate is given by eq.~(\ref{eq:nucoll}) after
substitution of $\Rmin$ from eq.~(\ref{eq:Rmintid}). We assume $f_1 =
f_\mathrm{NS}(t)$ and $f_2 = f_\mathrm{WD}(t)$ to vary during the
early-phase collapse from their primordial values to respectively
$0.547$ and $0.218$, and then to remain the same during the late-phase
collapse. This yields the expected number ($\simeq 8.86$) of tidally
captured NS-WD binaries with (pericenter) seperations up to
$0.109$\,\Rsun.
This is well below the upper limit of $1.669$\,\Rsun\ that increases
after the exchange interaction to the maximum separation of
$3.93$\,\Rsun\ allowed for merger of the dynamically formed NS-NS
binary within $\simeq 6$\,Gyr. The lower limit to avoid stable mass
transfer is $0.054$\,\Rsun, and gives rise to a fraction
$f_\mathrm{sep} \simeq 0.511$ of tidally captured NS-WD binaries with
separations in this allowed range, and with corresponding mean
separation of $\overline{a} \simeq 0.093$\,\Rsun. Finally, we divide
by the $\simeq 2.86 \times 10^3$ stellar systems in the core to arrive
at a fraction $f_\mathrm{NS-WD,tid} \simeq 1.72 \times 10^{-3}$ of
tidally captured NS-WD binaries.
The calculation for the subsequent exchange reaction with a NS, is the
same as above in Section~\ref{sec:fromprimordialbinaries} for the
formation of NS-NS binaries from primordial NS-WD binaries, except
that in eq.~(\ref{eq:nucoll}) we now have $f_1 =
f_\mathrm{NS-WD,tid}$.
In this way, we estimate that the rate $\Rdyn$ to turn a tidally
captured NS-WD binary dynamically into a NS-NS binary that can merge
within the next $\simeq 6$\,Gyr is a factor $\Rdyn/\Rtid \simeq 3.70$
larger than the tidal capture rate.

We now turn to the additional dynamical formation of a NS-NS binary
from a tidally captured WD-WD binary through two subsequent three-body
exchange interactions with a NS. Since the tidal capture rates 
between two WDs and between a WD and a NS are similar
(Table~\ref{tab:rates}), we expect the rate to be significantly less
than just estimated for NS-NS binary formation from a tidally captured
NS-WD binary, which only needs one subsequent exchange interaction.
We assume that both the tidal capture of the two WDs and the first
subsequent exchange happen before deepest core collapse.
This means we first calculate the fraction of WD-WD binaries formed
through tidal capture as function of time, i.e.,
$f_\mathrm{WD-WD,tid}(t)$. The calculation is the same as for
$f_\mathrm{NS-WD,tid}$ above, but with $f_1 = f_2 = f_\mathrm{WD}(t)$,
and we integrate the tidal capture rate up to time $t$, followed by
division with the number of stellar systems in the core $N_c(t) =
(4\pi/3) n_c(t) r_c(t)^3$ at that time. To have the WD-WD binary
(pericenter) seperations up to $0.092$\,\Rsun\ within the allowed
range of $[0.043,0.708]$\,\Rsun, implies $f_\mathrm{sep} \simeq 0.530$
and a mean separation of $\overline{a} \simeq 0.072$\,\Rsun.
Next, we substitute $\Rmin = \overline{a} + r_\mathrm{NS} \simeq
0.072$\,\Rsun, $f_1 = f_\mathrm{WD-WD,tid}(t)$, $f_2 =
f_\mathrm{NS}(t)$ and $m_1 + m_2 \simeq 2 \times 0.5 + 1.4 =
2.4$\,\Msun\ in eq.~(\ref{eq:nucoll}). Integrating the resulting rate
over the full duration of the core collapse phase, yields the expected
number ($\simeq 4.08 \times 10^{-3}$) of dynamically formed NS-WD
binaries from tidally captured WD-WD binaries. We then divide by the
$\simeq 2.86 \times 10^3$ stellar systems in the core at deepest
collapse, to arrive at a fraction $f_\mathrm{NS-WD,dyn} \simeq 1.43
\times 10^{-6}$.
As above in Section~\ref{sec:fromprimordialbinaries}, we repeat the
rate calculation for dynamically forming a NS-NS binary in a second
exchange interaction with a NS. In this way, we find that the rate to
form a NS-NS binary dynamically from a tidally captured WD-WD binary
through two subsequent three-body exchange interactions is smaller
than the tidal capture rate by a factor $\Rdyn/\Rtid \simeq 0.555
\times 10^{-2}$.

By combining the above two formation channels, we thus expect that the
rate to dynamically turn tidally captured binaries into compact NS-NS
binaries that can merge in time is larger than the tidal capture rate
by a factor $\simeq 3.71$. As expected, this factor is mainly the
result of the first formation channel, of which the rate is much
larger, by a factor $\simeq 667$, than the second formation channel.
Next, we estimate in a similar way also the rates to dynamically form
BH-NS binaries from tidally captured NS-WD or BH-WD binaries via a
single exchange interaction with respectively a BH or NS, or from
tidally captured WD-WD binaries via two subsequent exchanges
interaction with both a NS and a BH.
From the three formation channels together, we predict a formation
rate of BH-NS binaries that is larger than the tidal capture rate
between a single BH and NS by a factor $\simeq 8.42$. Again, the
contribution from the third formation channel that involves two
subsequent exchange interactions is much smaller, by factors $\simeq
72.8$ and $\simeq 371$, than the first and second formation channel.

\subsubsection{Merger versus encounter rate}
\label{sec:mergervsencounter}

The above calculations provide an estimate of the relative frequencies
of (binary) mergers and close encounters between two compact
objects in the cores of GCs.
Combining the results from
Sections~\ref{sec:primordialcompactbinaries} --
\ref{sec:fromtidallycapturedbinaries}, we estimate for a compact NS-NS
(BH-NS) binary a higher merger than tidal capture rate by a factor
$\simeq 3.7$ ($\simeq 8.4$).

These estimates can be further improved by allowing a range in WD and
MS star masses (instead of adopting mean masses of $0.5$\,\Msun\ and
$0.4$\,\Msun, respectively), by sampling the full distribution in the
ratio of binary separations before and after a three-body exchange
interaction (instead of taking the median value of
$a_\mathrm{fin}/a_\mathrm{ini}$), and by including the formation and
subsequent merger of compact binaries throughout the lifetime of a GC
instead of concentrating on deepest core collapse with a peak around
$z \simeq 0.7$).
Also the binaries will interact among themselves, so that four-body
interactions could become important, especially since the number of
tidally captured binaries is increasing; when for all ten possible
binary pairings from a BH, NS, WD and MS, we integrate the
corresponding tidal capture rate up to the time of deepest collapse,
we find a total of $\simeq 207$ dynamically formed binaries. Given the
$\simeq 2.86 \times 10^3$ stellar systems in the core, we obtain a
binary fraction of $\simeq 7.23$\,\% compared to a primordial binary
fraction of a few per cent.
These and other improvements are clearly interesting but beyond the
scope of this paper.

From the above estimates, we can conclude that in the cores of GCs the
rate of close encounters between two single compact objects can become
similar to the rate of merging of two compact objects in a binary.
The latter compact binaries most likely are formed through a
three-body exchange interaction with a binary which itself originated
from a tidal capture between a compact and non-compact (WD) stellar
object (Section~\ref{sec:fromtidallycapturedbinaries}). The
alternative dynamical formation of compact binaries from primordial
binaries (Section~\ref{sec:fromprimordialbinaries}) is less likely by
about two orders of magnitudes. A main reason is that the fraction of
primordial binaries in GCs is at most a few per cent, which in turn
also implies that the number of primordial compact binaries in GCs is
negligble, while they are thought to be the dominant source of SGRBs
in the ``field'' of a galaxy
(Section~\ref{sec:primordialcompactbinaries}).
While SGRBs are commonly assumed to orginated from compact binary
mergers, we thus find that close encounters between two compact
objects in the core of GCs can also provide a significant
contribution.

\section{Discussion}\label{sec:disc}

Dense stellar systems in some ways are like an ecological network,
where feedback is extremely important and apparently isolated events
can have far-reaching consequences. In a globular cluster, as we
discussed, dynamical interactions between passing stars can form new
binaries and modify the properties and even the membership of existing
binaries.  Motivated by this, we have investigated the production of
compact binaries via two and three body encounters.  We find that
event rates within globular clusters are expected to be significant,
and can become similar to the overall production of merging compact
binaries. This hints at the underlying possibility that SGRB
progenitors may not be entirely restricted to the most widely favored
scenario involving the merger of compact binaries in the field. Much
of our effort in this section will therefore be dedicated to
determining what are the expected characteristics of SGRBs arising
from these encounters and how do they compare to recent observational
constraints.

\subsection{Prospects for the production of SGRBs}\label{sec:SGRBs}

It is clear from the calculations presented in Section~\ref{sec:hydro} that
the formation of an accretion disk around the primary is a robust
result, regardless of the initial orbital parameters (as long as
disruption occurs, of course). The particulars of each case, however,
are variable in several aspects, and can lead to interesting
diversity.

First, and most importantly, the resulting disk mass is not always
equal. This is crucial in terms of the available energy in the system,
as it sets the overall scale for an accretion powered event that could
produce a SGRB. We find that even for the relatively small variation
in angular momentum (or equivalently, impact parameter) for black
hole--neutron star encounters the resulting disk mass, $M_{\rm disk}$,
varies almost by a factor of three. We have previously estimated
\citep{lrrg05,lrr07} that the energy that can be potentially extracted
from such disks to produce a GRB scales as $M_{\rm disk}^{2}$ for
neutrino powered events, and as $M_{\rm disk}$ for magnetically
dominated bursts. The range in mass thus possibly spans an order of
magnitude when converted to total energy release.

Second, the nature of the primary itself can produce a different
outcome. The ``cleanest'' scenario involving a black hole leads to
accretion in the hypercritical regime discussed above. However, for
double neutron star collisions, a range of possibilities remains. For
the one calculation of this type we have performed, at the end of the
simulation the central core is rapidly rotating and is surrounded by
a massive envelope. If the core can avoid collapse, it is possible
that the rapid rotation will wind up the magnetic field to large
magnetar-like values \citep{price06} and allow for repeated episodes of
energy release \citep{usov92,kluzniak98}.

Third, the presence of large tidal tails in which material will fall
back onto the central object at a later time is a generic feature of
the present set of calculations. The mass involved in these structures
is considerably larger than for binary mergers for the reasons already
mentioned in Section~\ref{sec:tails}. As the impact parameter in the black
hole neutron star collisions increases, the mass in the tail can
become even larger than that in the disk. Thus the properties of the
fall back material will dominate the behavior at late times
\citep{lrrlc09}.

\begin{figure}
  \begin{center}
    \includegraphics[width=0.4\textwidth]{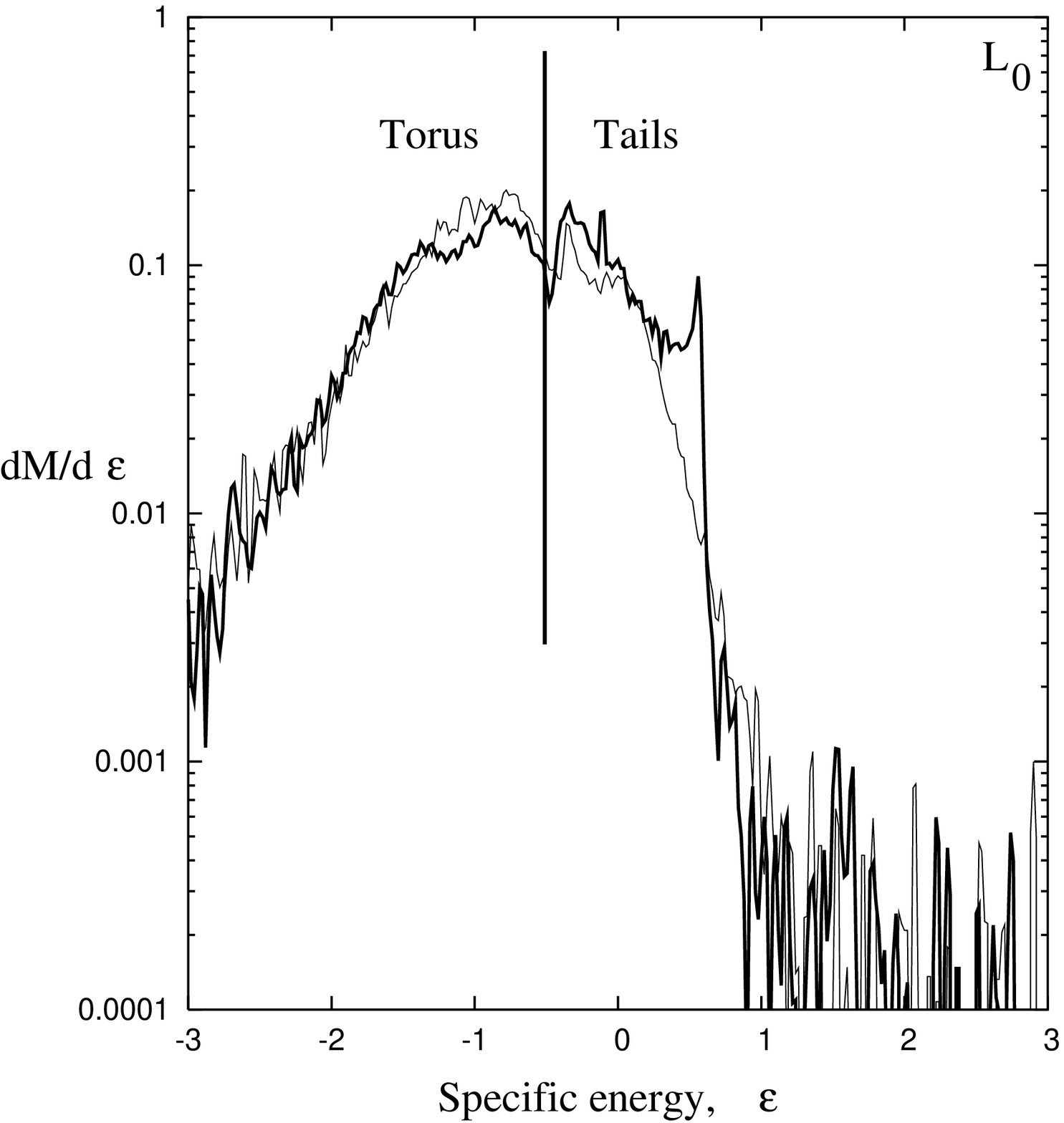}
    \hfill
    \includegraphics[width=0.4\textwidth]{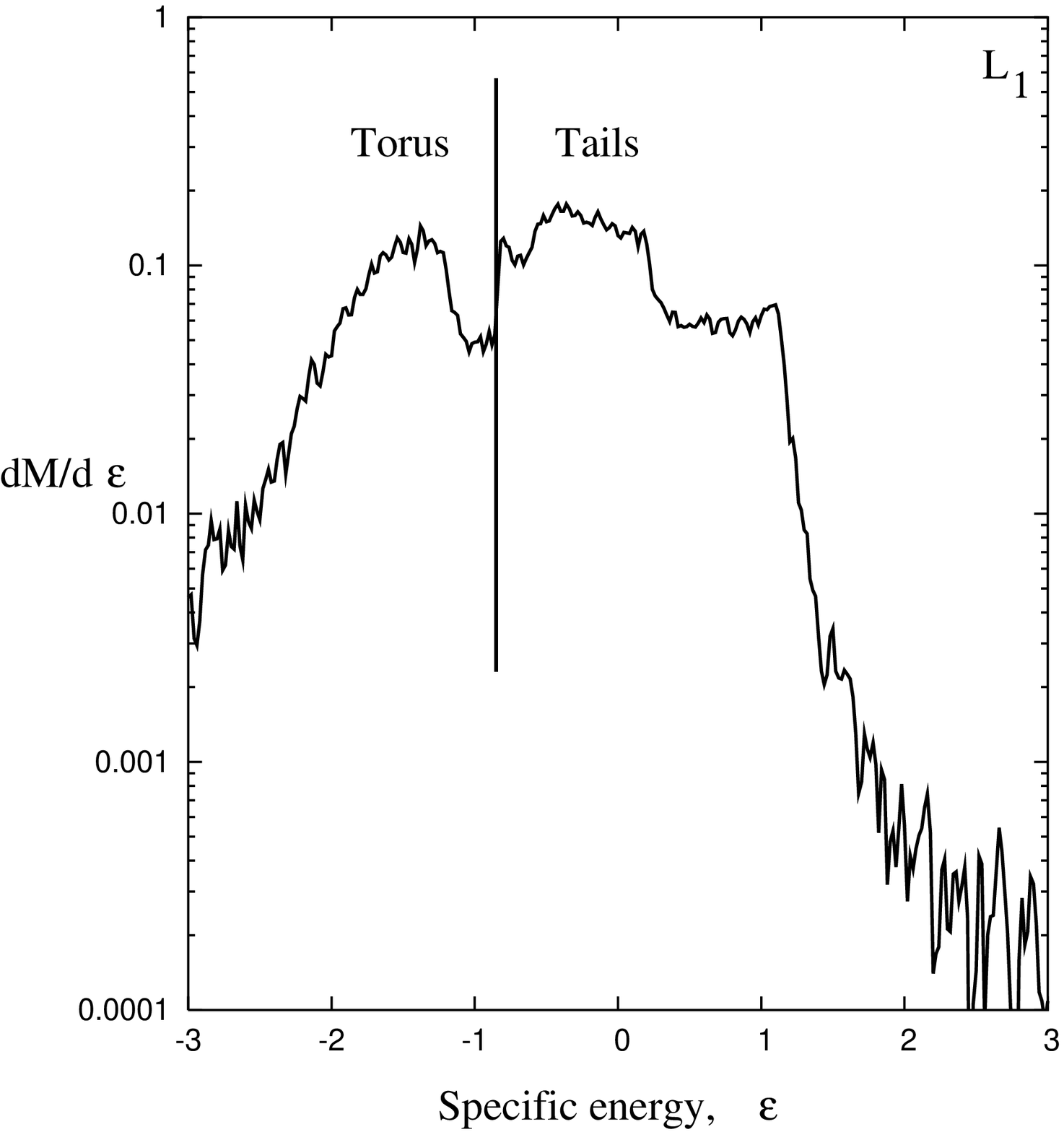}
     \hfill
    \includegraphics[width=0.4\textwidth]{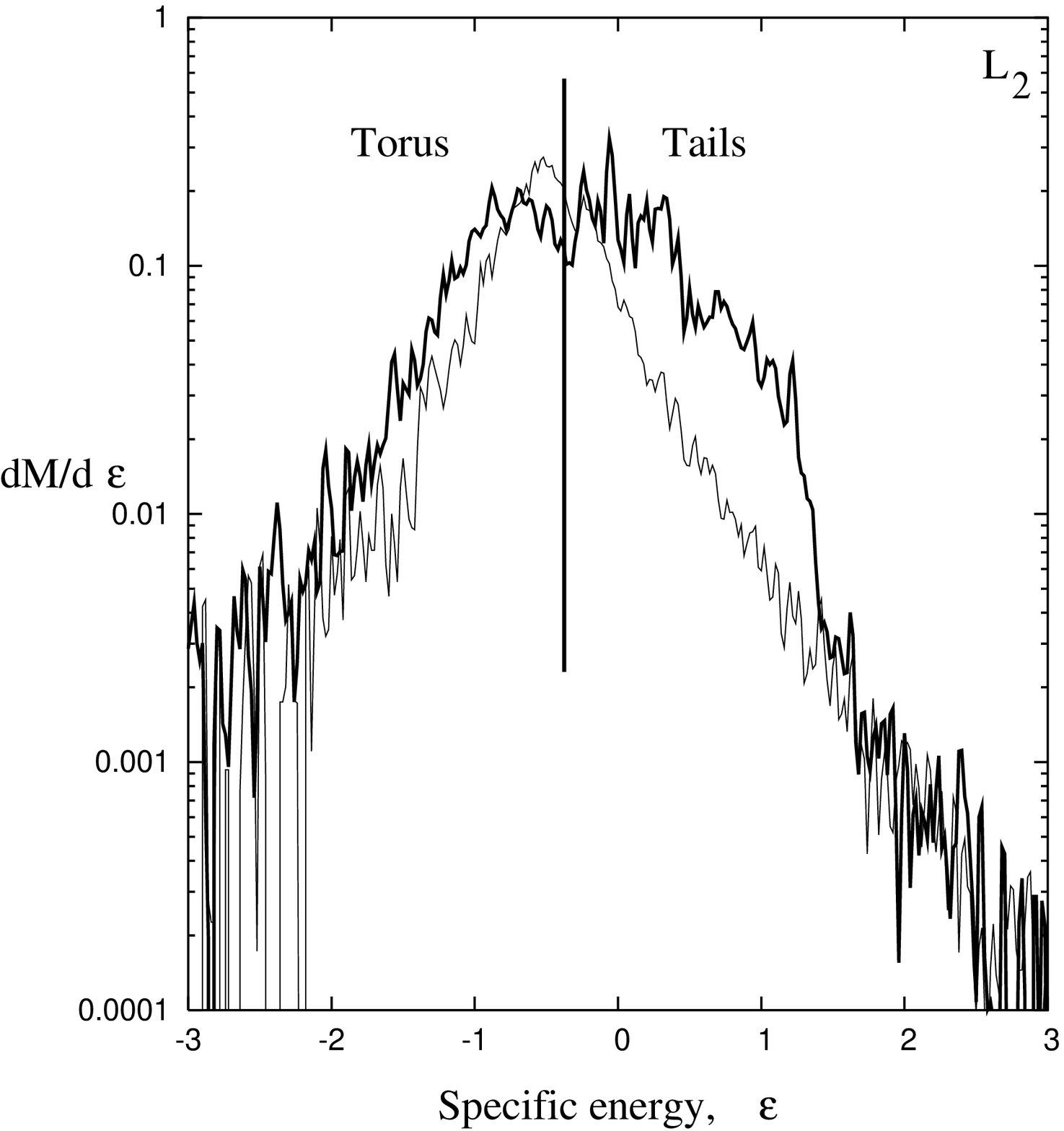}
 \end{center}
 \caption{Differential distribution of mass with specific energy for
   the fluid at the end of runs L$_{0}$, L$_{0}\Gamma_{5/3}$, L$_{1}$,
   L$_{2}$ and L$_{2}\Gamma_{5/3}$. The thick (thin) lines are for
   $\Gamma=2$ ($\Gamma=5/3$). The vertical line marks the
   morphological division between matter that lies in the torus around
   the black hole and that contained in the tidal tails. A fraction of
   the latter has positive energy and will escape the system.}
\label{fig:dmdeL}
\end{figure}

\begin{figure}
\includegraphics[width=\columnwidth,angle=0,scale=1.]{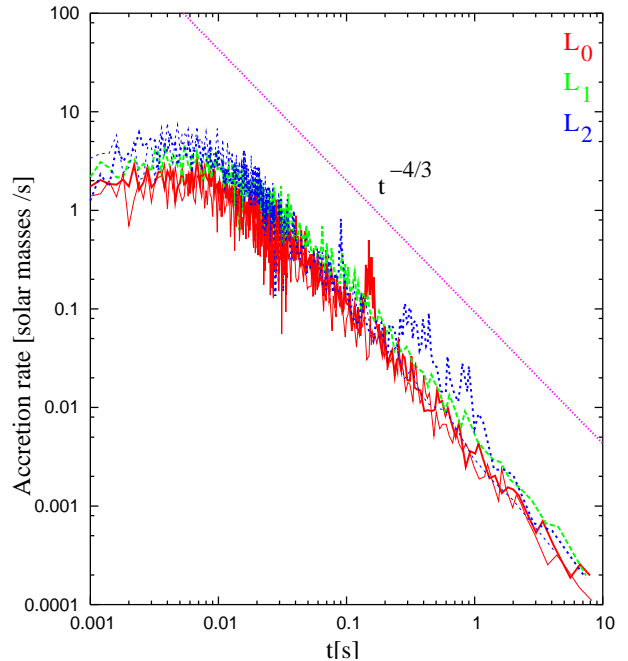}
\caption{The fall back accretion rate onto the central object from
  tidal tails (in solar masses per second) is shown for runs L$_{0}$,
  L$_{0}\Gamma_{5/3}$, L$_{1}$, L$_{2}$ and L$_{2}\Gamma_{5/3}$, along
  with a reference power law with decay index $-4/3$. The thick (thin)
  lines are for $\Gamma=2$ ($\Gamma=5/3$).}
\label{fig:mdotfall}
\end{figure}

The differential energy distribution in the tails is shown in
Figure~\ref{fig:dmdeL}. Material with negative energy is either in the
torus surrounding the black hole, or in a portion of the tail that is
bound to it. Fluid with positive energy will eventually leave the
system. The thick black vertical line in each plot separates the torus
and the tails morphologically, making it clear that a substantial
portion of the tails is bound and will fall back. In the simple
analytical estimates performed initially for stellar interactions with
supermassive black holes, the distribution of mass with specific
energy of the material from the disrupted star was constant, thus
giving rise to a fall back accretion rate $\dot{M}_{\rm fb} \propto
t^{-5/3}$, computed assuming ballistic trajectories \citep{rees88}. We
have computed the corresponding accretion rate here as well, also by
assuming that the fluid in the tails is on ballistic trajectories in
the potential well of the central mass (allowing us to follow it for a
much longer time). The more complex interaction we have outlined for
the case of comparable masses produces a different decay law, closer
to $\dot{M}_{\rm fb} \propto t^{-4/3}$, see
Figure~\ref{fig:mdotfall}. There is also variability on shorter time
scales superposed on this decay, due to inhomogeneities in the tidal
tails. Characterizing this requires a full hydrodynamical and
thermodynamical treatment of the motion in the tails at late times
\citep{rosswog07,lrrlc09,metzger09}.

Essentially, the bulk of the material in the tail will return to the
vicinity of the compact object within a few seconds. As it has finite
angular momentum, it will not directly impact the primary but describe
an accentric orbit around it. Moreover, since it is not composed of
test particles but is a fluid stream, dissipation close to periastron
will circularize the orbit at a radius roughly equal to the distance
from which it was ejected in the first place. If angular momentum
transport processes are present and the gas is able to cool with some
efficiency, it will then form an accretion disk and feed the primary
with mass and energy. The viscous time scale will be much longer than
the dynamical re-injection time scale, and thus for transport purposes
it will be as if a disk had been formed practically impulsively around
the black hole (or neutron star), subsequently evolving on a secular
time scale. It is thus possible, in principle, to account generically
in this scenario for a prompt episode of energy release as well as for
activity many dynamical time scales later.

Finally, we have shown also that the nature of the secondary (neutron
star vs. white dwarf) will lead to a substantially different final
configuration. In the white dwarf case the accretion disk is much
larger, and thus the densities are substantially lower (the total mass
being comparable). The key question in this case is whether the gas
can find an efficient mechanism to cool, and thus accrete. Otherwise
the material will only be blown to large radii and not release enough
gravitational binding energy to account for the energy budget of a
GRB.

\subsection{Time delay}
\label{sec:timedelay}

The observed event rate of SGRBs as function of redshift can provide
constraints on a time delay between the formation of the progenitors
and the explosion \citep{guetta05, guetta06, nakar06,
  bloomprochaska06}.
SGRBs exhibit great diversity in terms of their host galaxies, and a cursory comparison of the redshift distribution of SGRBs with the
universal star formation rate (SFR) reveals what appears to be a
significant time delay of a few Gyr.  A large progenitor lifetime would help explain the
apparent high incidence of galaxy cluster membership
\citep{Ped05,bloom06,berger07b}, while on  the other hand, shorter lifetimes
are required to explain the population of SGRBs at moderately high
redshift \citep{berger07a,graham09}. Short delays have been pointed out as a possibility  for merging binaries directly from populations synthesis calculations \citep{belczynski02,belczynski06,osh08}, and could be responsible for a significant fraction of the observed event rate. 

An alternative approach for constraining the distribution of time delays may be to use the event
rates of SGRBs in different types of galaxies \citep{gal-yam08,zheng07}. On average, early-type galaxies have their stars formed earlier than late-type galaxies, and this difference, together with the time delay between progenitor formation and SGRB outburst, inevitably leads to different burst rates in the two types of galaxies. For instance, the morphological types for SGRBs reflect a higher incidence of early-type galaxies than Type Ia supernovae and this would suggest associated progenitor lifetimes significantly exceeding a few Gyrs \citep{zheng07}.  
Making more quantitative statements about the ages of the progenitor
systems is not only hampered by small number statistics but also from
the lack of robust predictions of the distribution of merger sites as
a function of time.

\begin{figure}
\includegraphics[width=\columnwidth,angle=0,scale=1.]{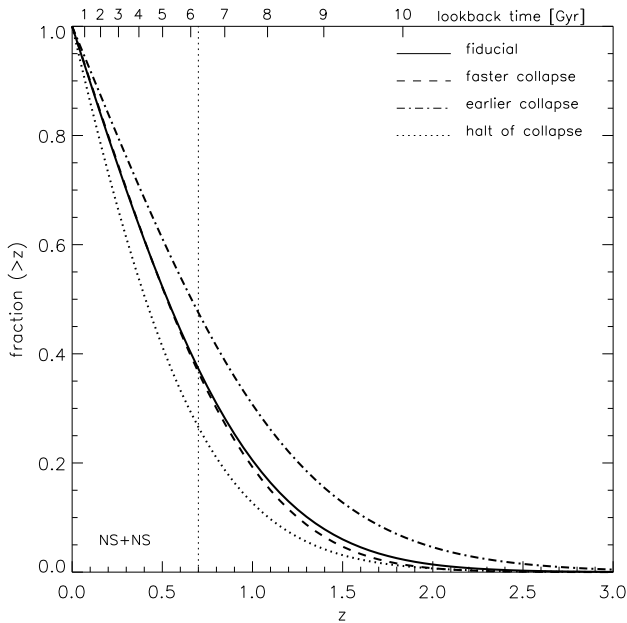}
\caption{The predicted cumulative redshift distribution of SGRBs under
  the assumption that they are produced by tidal capture of two
  neutron stars in the cores of globular clusters. The meaning of
  different line styles is the same as in Fig.\ref{fig:m15_nuencevo}.}
\label{fig:m15_nuenccumul}
\end{figure}

Observational evidence for a significant time delay between tracers of
star formation and SGRB outburst excludes that the production of SGRBs
is exclusively related to the short time-scale for evolution and death of
massive stars, which are believed to be the progenitors for long GRBs.
However, significant time lags between the cosmic star formation rate
and the SGRB redshift distribution occur naturally if the progenitor is a
compact binary in the field that merges \citep{cenko08,salvaterra08,hopman06}, as
well as in the case of a close encounter between two compact stellar objects in a dense stellar environment. For a merger, the time delay reflects the time for the two compact objects
to merge by emission of gravitational waves, while for close encounters
a significant rate is only achieved when the stellar density rises significantly through the core collapse of a stellar cluster. Figure~\ref{fig:m15_nuenccumul} shows our predicted
cumulative redshift distribution for the progenitors of SGRBs assuming
they are produced by tidal capture of two neutron stars in GCs.
Further redshift determinations are required to help differentiate
between various ways of forming a short GRB, although, as we argue
next in Section~\ref{sec:galacticenvironment}, detailed observations
of the astrophysics of individual GRB host galaxies may be essential
before stringent constraints on the lifetime of short GRB progenitors
can be placed.

\subsection{Galactic environment}
\label{sec:galacticenvironment}

The observed offsets from what has been argued are the plausible
hosts, if true, also holds important ramifications for the sort of
viable progenitors \citep[see][]{belczynski06,zemp09}.  Very large offsets seen from early-type hosts
would seem to be at odds with progenitor systems with small systematic
kicks such as those occurring in globular clusters, although with such
large physical offsets the possibility remains that the association
with the putative host is coincidental. On the other hand, based on
the small offsets from some low-mass galaxy hosts
\citep{prochaska06,soderberg06,berger09,bloomprochaska06}, SGRB
progenitors cannot all have large systematic kicks at birth and
inherently large delay times from formation.

Compact binaries in the field are expected to experience a kick,
leading to mergers away from their point of origin. The displacement
depends on the distribution of kick velocities, merger times, and host
masses, with predicted values in the range 10--100\,kpc
\citep{fryer99, belczynski06,zemp09}, depending on the formation channel, host galaxy type and mass.  By contrast, GCs are expected to show
on average moderate displacements: the spatial distribution of GCs
peaks around the half-light radius ---typically a few kpc--- of the
host galaxy.  What is more, due to the large red giant star density in
the GC core, there is the possibility for the interaction of the
external shock with a denser external medium than that of the IGM
\citep{prrl09}.

If a significant fraction of SGRBs is indeed produced in GCs we also
expect to see a strong preference for this scenario occurring in luminous host galaxies, because
the number of GCs increases steeply with the host galaxy luminosity as
$\propto L_V^{>1.5}$ \citep{mclaughlin99}, whereas the number of
primordial binaries increases only as $\propto L_V$. Red,
early-type galaxies are on average more luminous than blue, late-type
galaxies, which would lead to  a higher incidence of this kind of SGRB in ellipticals when compared to spiral galaxies. The current limited sample indicates that, if unidentified hosts are in fact ellipticals, SGRBs would be approximately evenly distributed between early and late-type hosts \citep{berger09}. However, the morphology of galaxies changes with
time due to internal (secular) evolution, and in particular when they
merge. As a result, the fraction of late-type to early-type galaxies
increases toward higher redshift, and hence SGRBs in GCs might occur
also more frequently in spiral galaxies.

In other words, if (old) GCs are formed in the highest density peaks
before reionization \citep[e.g.][]{moore06}, and subsequently were
accreted in galaxies through mergers, SGRBs or any other transient
connected with GCs would provide a unique tracer of the hierchical
build-up of galaxies. This look-back on galaxy evolution is of course
limited by the observability of such transients, both due to their
intrinsic brightness and the duration of the event, as well as due to
the time delay discussed in Section~\ref{sec:timedelay}. The latter
time delay is probably also what limits the possible contribution from
intermediate-age GCs which are thought to form in gas-rich galaxy
mergers. The formation time of these intermediate-age GCs plus the
time for their cores to collapse will significantly shift the
potential production of SGRBs towards lower redshift with respect to
old GCs.

Up to this point, we have argued for the production of SGRBs in GCs as
oppposed to originating from primordial compact binaries within the
host galaxy field, but we have not yet distinguised between mergers or
close encounters between compact stellar objects in GCs. Like
intermediate-age versus old GCs, we also expect the redshift
distribution of SGRBs in the merger scenario to be skewed toward lower
redshift than in the encounter scenario. From the global encounter
rate in Figure~\ref{fig:m15_nuencspace}, we expect a significant SGRB
event rate only below $z \sim 1$. This is still consistent with SGRBs
found recently around $z \sim 1$ \citep{berger07a}, but is very
challenging or perhaps impossible in case of the additional time delay
in case of merging compact binaries. 

\subsection{Nuclear clusters}
\label{sec:nuclearclusters}

The high stellar densities in which both close encounters and
three-body interactions become significant may, besides in the
collapsed cores of GCs, also be reached in the nuclei of galaxies.  Of
particular interest are the nuclear (stellar) clusters (NCs) as they
might also experience core collapse.
NCs have been found in 50\%--75\% of both late-type and early-type
galaxies, but they are absent in the elliptical galaxies brighter than
absolute magnitude $M_B \sim -19$ (although this could in part be an
observational bias due to the presence of steep central cusps), and
the frequency of nucleation also falls to zero in dwarf galaxies
fainter than $M_B \sim -12$ \citep[see the review by][]{boker08}.
NCs have typical half-light radii \citep[$\sim 3.5$\,pc;][]{boker02}
similar to GCs, but their typical masses \citep[$\sim 3 \times
10^6$\,\Msun;][]{walcher05}, and hence their average density is also
higher than that of GCs by the same factor.

On the other hand, the half-mass relaxation time of NCs is about an
order of magnitude longer than those in GCs, both due to their higher
typical mass as well as due to the higher velocity dispersion of the
surrounding galaxy. The latter results in a flow of energy from the
galaxy to the NC, which opposes core collapse and might even lead to
core expansion \citep{dokuchaev85, kandrup90}. The result is a minimum
compactness of NCs in order to resist expansion \citep{quinlan96,
  merritt09}, which is close to their observed sizes.  This, together
with their longer half-mass relaxation times when compared with GCs,
argues against NCs experiencing core collapse and thus providing a
fertile ground for the production of SGRBs via close encounters
between compact stellar objects. In addition, the presence of a
possible (super or intermediate) massive black halo in NCs
\citep{seth08} inhibts core collapse, and NCs with black holes always
expand \citep{merritt09}.

Still, it is possible that NCs were formed with a high initial
concentration (and without an as yet  significantly massive black hole) at
the highest density peaks so that their cores collapsed early on.
Later on the core and NC as a whole expanded, partly at least due to
growing heat input from the surrounding galaxy of which the velocity
dispersion increases as a result of its (secular and/or hierarchical)
evolution.
An additional and likely even stronger expansion of a NC is expected
in case the host (dwarf) galaxy is stripped away when it is accreted
by a larger galaxy; a formation history that is believed to be
applicable also to the most massive GCs in the Milky
\citep[e.g][]{georgiev09}.
In both cases, the currently observed larger size and longer half-mass
relaxation time would not be an indication of a possible core collapse in
the past.

If NCs indeed experienced such a core collapse significantly earlier
than GCs, we expect from comparing the earlier collapse and fiducial
model in Figure~\ref{fig:m15_nuencspace} that the SGRBs potentially
produced through close encounters of compact objects in NCs will be
skewed to higher redshift $z>0.7$. At the same time, we expect the
SGRB to be located in the center of typically a faint (nucleated)
dwarf galaxy. Also, the multiple stellar populations observed in local
nuclear clusters \citep[e.g.][]{walcher06} very likely imply on-going
star formation at higher redshift. This is all in line with the
recently observed high-redshift SGRBs and their host galaxy properties
\citep{berger09}, but clearly requires further study of NCs and in
particular evidence for past core collapse.

We finally note that even though the presence of a central black hole
inhibits core collapse and hence prevents short (two-body) relaxation
times, secular resonant relaxation can operate on a much smaller time
scale. This alternative relaxation process can increase the
ellipticity of stellar objects on (near) Keplerian orbits, and hence
not only bring them closer to the black hole \citep[e.g.][]{rauch96},
but also to each other. This increases again the change of close
encounters, and, if it involves two compact objects, the potential to
produce SGRBs in the vicinity of massive black holes.

\section{Summary and conclusions}\label{sec:ccl}
It is evident from the work presented here that tidal capture in close encounters and collisions
between compact stellar objects in the cores of GCs can provide a
viable channel for the production of SGRBs at rates which are significant when compared to those from the
mergers of primordial compact binaries, in both GCs and the field. For GCs with small primordial binary populations, we have argued that the formation of coalescing compact binaries which are capable of powering a SGRB is dominated by dynamical exchanges of NSs (BHs) with tidally formed WD-NS(BH) binaries and by the close encounters of NSs with NS/BHs. 

Through a realistic though conservative calculation, we predict an event rate for the tidal capture and collision of two neutron stars in the collapsed core of a GC 
that steeply increases to $\sim 50$\,\pyrGpcc\ around $z \sim 0.7$, and is followed by a gradual decline to $\sim 30$\,\pyrGpcc\ at $z = 0$. This is consistent with the currently
observed event rate and redshift distribution of SGRBs. Furthermore,
since the number of GCs both steeply increases with galaxy luminosity
and peaks at the half-light radius of the host galaxy, we expect SGRBs
to appear in both late and early-type galaxies and to be displaced
from the galaxy center, in line with the cursory identification of and
location in host galaxies.

Using detailed hydrodynamics simulations we have explored the collision of NSs with WD/NS/BHs for a range of system parameters, to complement previous knowledge on binary mergers. We have shown that close encounters involving neutron stars and/or black holes can account for both a prompt episode of energy release, as well as for activity many dynamical times later through a accretion disk formed from the re-infall of disrupted material expelled in a bound tail. Relatively small changes in the impact parameter of the close encounter, as well as the nature of the
secondary clearly result in a significant diversity in the final
configuration. The significant amount of material that is dynamically ejected comes from the multiple periastron passages experienced before full tidal disruption takes place, each one carrying a distinctive signature of its thermodynamical history. Thus, while it could contribute significantly to the observed abundance of r-process elements \citep{ls74,ls76,symbalisty82,frei99}, not all of it need do so in principle.  

\acknowledgments We thank J. Bloom, C. Fryer, J. Guillochon,
D. Heggie, P. Hut, V. Kalogera, R. O'Shaughnessy, X. Prochaska,
F. Rasio and S. Rosswog for useful discussions and comments on the
draft.
We thank both referees for constructive comments on this work, which helped improve the final version.
Part of this work was carried out during visits to
the Institute for Advanced Study in Princeton, the University of
California in Santa Cruz, and the Instituto de Astronom\'{\i}a at
UNAM, whose hospitatlity is gratefully acknowledged. This work was
supported in part by CONACyT-83254 and DGAPA-UNAM-IN-113007 (WL), NASA
NNX08AN88G and the David and Lucile Packard Foundation (ER), UCMEXUS
(ER and WL) and NASA through Hubble Fellowship grant
HST-HF-01202.01-A, awarded by the Space Telescope Science Institute,
which is operated by the Association of Universities for Research in
Astronomy, Inc., for NASA, under contract NAS 5-26555 (GvdV).


\begin{thebibliography}{}

\bibitem[Backer et al.(1982)]{backer82} Backer, D.C., Kulkarni, S.R.,
  Heiles, C., Davis, M.M., Goss, W.M. 1982, Nature, 300, 615

\bibitem[Barthelmy et al.(2005)]{barthelmy05} Barthelmy,
  S.~D. et~al. 2005, Nature, 438, 994

\bibitem[{{Baumgardt} {et~al.}(2003){Baumgardt}, {Heggie}, {Hut}, \&
  {Makino}}]{baumgardt03}
{Baumgardt}, H., {Heggie}, D.~C., {Hut}, P., \& {Makino}, J. 2003, \mnras, 341,
  247

\bibitem[Baumgarte et al.(2000)]{baumgarte00} Baumgarte, T.W.,
  Shapiro, S.L., Shibata, M. 2000, \apj, 528 L28

\bibitem[Belczynski, Bulik \& Kalogera(2002)]{belczynski02} Belczynski, K., Bulik, T., Kalogera, T. 2002, \apj, 571, L147

\bibitem[{Belczynski} {et~al.}(2006)]{belczynski06}
{Belczynski}, K., {Perna}, R., {Bulik}, T., {Kalogera}, V., {Ivanova}, N., \&
  {Lamb}, D.~Q. 2006, \apj, 648, 1110

\bibitem[Berger(2007)]{berger07b} Berger, E. 2007, \apj, 670, 1254

\bibitem[{Berger}(2009)]{berger09} {Berger}, E. 2009, \apj, 690, 231

\bibitem[Berger et al.(2005)]{berger05} Berger, E. et al. 2005,
  Nature, 438, 988

\bibitem[Berger et~al.(2007)]{berger07a} Berger, E. et al. 2007, \apj,
  664, 1000

\bibitem[{Bettwieser} \& {Sugimoto}(1984)]{bettwieser84}
{Bettwieser}, E. \& {Sugimoto}, D. 1984, \mnras, 208, 493

\bibitem[Bloom et al.(2006)]{bloom06} Bloom, J.~S. et al. 2006, ApJ,
  638, 354

\bibitem[Bloom \& Prochaska(2006)]{bloomprochaska06} Bloom, J.~S.,
  Prochaska, J.~X. 2006, AIP Conference Series, 836, 473
  
\bibitem[{B{\"o}ker}(2008)]{boker08}
{B{\"o}ker}, T. 2008, Journal of Physics Conference Series, 131, 012043

\bibitem[{B{\"o}ker} {et~al.}(2002)]{boker02} {B{\"o}ker}, T.,
  {Laine}, S., {van der Marel}, R.~P., {Sarzi}, M., {Rix}, H.-W.,
  {Ho}, L.~C., \& {Shields}, J.~C. 2002, \aj, 123, 1389


\bibitem[Carter \& Luminet(1983)]{carter83} Carter, B., Luminet,
  J.P. 1983, \aa, 121, 97
  
\bibitem[{Cenko} {et~al.}(2008)]{cenko08} {Cenko}, S.~B., {Berger},
  E., {Nakar}, E., {Kasliwal}, M.~M., {Cucchiara}, A., {Kulkarni},
  S.~R., {Ofek}, E., {Fox}, D.~B., {Harrison}, F.~A., {Rau}, A.,
  {Price}, P.~A., {Gal-Yam}, A., {Dopita}, M.~A., \& {Penprase}, B.~E.
  2008, \apj Letters submitted, arXiv:0802.0874v1

\bibitem[Chandrasekhar(1961)]{chandra61} Chandrasekhar, S. 1961, {\it
  Hydrodynamic and hydromagnetic stability} (Oxford: Oxford University
  Press)

\bibitem[{{Chernoff}(1993)}]{chernoff93}
{Chernoff}, D.~F. 1993, in ASP Conf. Ser. 50: Structure and Dynamics of
  Globular Clusters, ed. S.~G. {Djorgovski} \& G.~{Meylan}, 245--+

\bibitem[Clark(1975)]{clark75} Clark, G. W. 1975, ApJ, 199, L143

\bibitem[Cook et al.(1994)]{cook94} Cook, G.B., Shapiro, S.L.,
  Teukolsky, S.A. 1994, \apj, 424, 823
  
\bibitem[{Davis} {et~al.}(2008)]{davis08} {Davis}, D.~S., {Richer},
  H.~B., {Anderson}, J., {Brewer}, J., {Hurley}, J., {Kalirai}, J.~S.,
  {Rich}, R.~M., \& {Stetson}, P.~B. 2008, \aj, 135, 2155

\bibitem[{{De Marchi} {et~al.}(2007){De Marchi}, {Paresce}, \&
  {Pulone}}]{demarchi2007} {De Marchi}, G., {Paresce}, F., \&
  {Pulone}, L. 2007, \apjl, 656, L65

\bibitem[{Dokuchaev} \& {Ozernoi}(1985)]{dokuchaev85} {Dokuchaev},
  V.~I. \& {Ozernoi}, L.~M. 1985, Soviet Astronomy Letters, 11, 139

\bibitem[{Duquennoy} \& {Mayor}(1991)]{duquennoy91} {Duquennoy}, A. \&
  {Mayor}, M. 1991, \aap, 248, 485

\bibitem[Dull et al.(1997)]{dull97} Dull, J. D., Cohn, H. N., Lugger,
  P. M., Murphy, B. W., Seitzer, P. O., Callanan, P. J., Rutten,
  R. G. M. \& Charles, P. A. 1997, \apj, 481, 267

\bibitem[Eichler(1989)]{eichler89} Eichler, D. et al. 1989, Nature,
  340, 126

\bibitem[Fabian et al.(1975)]{fabian75} Fabian, A. C., Pringle,
  J. E. \& Rees, M. J. 1975, MNRAS, 172, 15

\bibitem[Faber et al.(2002)]{faber02} Faber, J.A., Grandcl\'{e}ment,
  P., Rasio, F.A. 2002, Phys. Rev. Letters, 89, 231102

\bibitem[Faber et al.(2007)]{faber07} Faber, S. M. et al. 2007, \apj,
  665, 265

\bibitem[Fox et al.(2005)]{fox05} Fox, D.~B. et~al. 2005, Nature, 437,
  845

\bibitem[Frank(1978)]{frank78} Frank, J. 1978, \mnras, 184, 87

\bibitem[{{Fregeau}(2008)}]{fregeau2008} {Fregeau}, J.~M. 2008, \apjl,
673, L25

\bibitem[Freiburghaus et al.(1999)]{frei99}Freiburgaus, C., Rosswog, S., Thielemann, F.-K. 1999, \apj, 525, L121

\bibitem[{Fryer} {et~al.}(1999)]{fryer99}
{Fryer}, C.~L., {Woosley}, S.~E., \& {Hartmann}, D.~H. 1999, \apj, 526, 152

\bibitem[Gal-Yam et al.(2008)]{gal-yam08} Gal-Yam, A. et al., 2008,
  \apj, 686, 408

\bibitem[Gehrels et al.(2005)]{gehrels05} Gehrels, N. et~al. 2005,
  Nature, 437, 851

\bibitem[Gehrels, Ramirez-Ruiz \& Fox(2009)]{gehrels09}Gehrels, N., Ramirez-Ruiz, E., Fox, D.B. 2009, ARA\&A, 47, 567

\bibitem[Graham et al.(2009)]{graham09} Graham, J.~F., et al.\ 
2009, \apj, 698, 1620

\bibitem[{{Gnedin} {et~al.}(1999){Gnedin}, {Lee}, \& {Ostriker}}]{gnedin99}
{Gnedin}, O.~Y., {Lee}, H.~M., \& {Ostriker}, J.~P. 1999, \apj, 522, 935

\bibitem[{Georgiev} {et~al.}(2009)]{georgiev09} {Georgiev}, I.~Y.,
  {Hilker}, M., {Puzia}, T.~H., {Goudfrooij}, P., \& {Baumgardt}, H.
  2009, \mnras, 647

\bibitem[{{Giersz} \& {Heggie}(2009)}]{gierszheggie2009}
{Giersz}, M. \& {Heggie}, D.~C. 2009, \mnras, 395, 1173

\bibitem[{{Goodman}(1987)}]{goodman87}
{Goodman}, J. 1987, \apj, 313, 576

\bibitem[Grindlay et al.(2006)]{grindlay06} Grindlay, J., Portegies
  Zwart, S. \& McMillan, S. 2006, Nature Physics, 2, 116

\bibitem[Guetta \& Piran(2005)]{guetta05} Guetta, D., Piran, T. 2005,
  A\&A, 435, 421

\bibitem[Guetta \& Piran(2006)]{guetta06} Guetta, D. \& Piran,
  T. 2006, A\&A, 453, 823

\bibitem[Hansen \& Murali(1998)]{hansen98} Hansen, B.M.S., Murali, C. 1998, \apj, 505, L15

\bibitem[Heggie(1975)]{heggie75} Heggie, D. C. 1975, MNRAS, 173, 72
  
\bibitem[{Heggie}(1985)]{heggie85} {Heggie}, D.~C. 1985, in IAU
  Symposium, Vol. 113, Dynamics of Star Clusters, ed. J.~{Goodman} \&
  P.~{Hut}, 139--157
  
\bibitem[{{Heggie} \& {Giersz}(2008)}]{heggiegiersz2008}
{Heggie}, D.~C. \& {Giersz}, M. 2008, \mnras, 389, 1858

\bibitem[{{Heggie} \& {Giersz}(2009)}]{heggiegiersz2009}
---. 2009, \mnras, 397, L46

\bibitem[{Heggie} \& {Ramamani}(1989)]{heggie89} {Heggie}, D.~C. \&
  {Ramamani}, N. 1989, \mnras, 237, 757

\bibitem[{Heggie} {et~al.}(2006)]{heggie06} {Heggie}, D.~C., {Trenti},
  M., \& {Hut}, P. 2006, \mnras, 368, 677

\bibitem[Heinke et al.(2006)]{heinke06} Heinke, C., et al., 2006, ApJ,
  651, 1098

\bibitem[{{H{\'e}non}(1965)}]{henon65}
{H{\'e}non}, M. 1965, Annales d'Astrophysique, 28, 62

\bibitem[Hessels et al.(2006)]{hessels06} Hessels, J. W. T., Ransom,
  S.  M., Stairs, I. H., Freire, P. C. C.,Kaspi, V.M., Camilo,
  F. 2006, Science, 311, 1901

\bibitem[Hills(1975)]{hills75} Hills, J. G. 1975, AJ, 80, 809

\bibitem[Hjorth et al.(2005)]{hjorth05} Hjorth, J. et al. 2005,
  Nature, 437, 859
  
\bibitem[{Hopkins} \& {Beacom}(2006)]{hopkins06} {Hopkins}, A.~M. \&
  {Beacom}, J.~F. 2006, \apj, 651, 142

\bibitem[{Hopman} {et~al.}(2006)]{hopman06} {Hopman}, C., {Guetta},
  D., {Waxman}, E., \& {Portegies Zwart}, S. 2006, \apjl, 643, L91

\bibitem[{Hurley} {et~al.}(2007)]{hurley07} {Hurley}, J.~R.,
  {Aarseth}, S.~J., \& {Shara}, M.~M. 2007, \apj, 665, 707
  
\bibitem[Hut \& Verbunt(1983)]{hut83} Hut, P., \& Verbunt, F.\ 1983,
  Nature, 301, 587
  
\bibitem[{Hut} {et~al.}(1992)]{hut92} {Hut}, P., {McMillan}, S.,
  {Goodman}, J., {Mateo}, M., {Phinney}, E.~S., {Pryor}, C., {Richer},
  H.~B., {Verbunt}, F., \& {Weinberg}, M. 1992, \pasp, 104, 981

\bibitem[{{Inagaki} \& {Lynden-Bell}(1983)}]{ilb83}
{Inagaki}, S. \& {Lynden-Bell}, D. 1983, \mnras, 205, 913

\bibitem[{Ivanova} {et~al.}(2008)]{ivanova08} {Ivanova}, N., {Heinke},
  C.~O., {Rasio}, F.~A., {Belczynski}, K., \& {Fregeau}, J.~M. 2008,
  \mnras, 386, 553

\bibitem[Janka \& Ruffert(1996)]{janka96b} Janka, T. \& Ruffert,
  M. 1996, A\&A, 307, L33

\bibitem[Jord\'{a}n et al.(2005)]{jordan05} {Jord{\'a}n}, A. et al. 2005,
  \apj, 634, 1002

\bibitem[Jord\'{a}n et al.(2006)]{jordan06} {Jord{\'a}n}, A. et
  al. 2006, \apjl, 651, L25

\bibitem[{Kalogera} {et~al.}(2004)]{kalogera04} {Kalogera}, V., {Kim},
  C., {Lorimer}, D.~R., {Burgay}, M., {D'Amico}, N., {Possenti}, A.,
  {Manchester}, R.~N., {Lyne}, A.~G., {Joshi}, B.~C., {McLaughlin},
  M.~A., {Kramer}, M., {Sarkissian}, J.~M., \& {Camilo}, F. 2004,
  \apjl, 601, L179

\bibitem[{Kandrup}(1990)]{kandrup90} {Kandrup}, H.~E. 1990, \apj, 364,
  100

\bibitem[Katz(1975)]{katz75}Katz, J. I. 1975, Nature, 253, 698

\bibitem[Kim \& Lee(1999)]{kimlee99}Kim, S.~S. \& Lee, H.~M. 1999,
  A\&A, 347, 123

\bibitem[Kochanek(1992)]{koch92} Kochanek, C.~S.\ 1992, \apj, 385, 604

\bibitem[Klu\'{z}niak \& Ruderman(1998)]{kluzniak98}Klu\'{z}niak, W.
  Ruderman, M. 1998, ApJ,

\bibitem[Lacy et al.(1982)]{lacy82}Lacy, J.H., Townes, C.H.,
Hollenbach, D.K. 1982, \apj, 262, 120

\bibitem[Lai et al.(1993a)]{lai93a}Lai, D., Rasio, F.A., Shapiro
S.L. 1993a, \apj, 406 L63

\bibitem[Lai et al.(1993b)]{lai93b}Lai, D., Rasio, F.A., Shapiro S.L. 1993b,
\apjs, 88, 205

\bibitem[Lattimer \& Schramm(1974)]{ls74}Lattimer, J.M., Schramm, D.N. 1974, \apj, 192, L145

\bibitem[Lattimer \& Schramm(1976)]{ls76}Lattimer, J.M., Schramm, D.N. 1976, \apj, 210, 549

\bibitem[Lee \& Ostriker(1986)]{lee86}Lee, H.~M., \& Ostriker,
  J.~P. 1986, \apj, 310, 176

\bibitem[Lee \& Kluzniak(1999)]{lee99a}Lee, W.H., Kluzniak, W. 1999,
\apj, 526, 178

\bibitem[Lee, Ramirez-Ruiz \& Granot(2005)]{lrrg05}Lee, W.H.,
  Ramirez-Ruiz, E., Granot, J. 2005, ApJ, 630, L165

\bibitem[Lee \& Ramirez-Ruiz(2007)]{lrr07}Lee, W.H. Ramirez-Ruiz, E. 
2007, New. J. Phys. 9. 17

\bibitem[Lee, Ramirez-Ruiz \& Lopez-Camara(2009)]{lrrlc09} Lee, W. H.,
  Ramirez-Ruiz, E. \& Lopez-Camara, D. 2009, \apj, 699, L93

\bibitem[Lee(2000)]{lee00} Lee, W.H. 2000, \mnras, 318, 606

\bibitem[Lee(2001)]{lee01} Lee, W.H. 2001, \mnras, 328, 583

\bibitem[{{Lynden-Bell} \& {Eggleton}(1980)}]{lbe80}
{Lynden-Bell}, D. \& {Eggleton}, P.~P. 1980, \mnras, 191, 483

\bibitem[McLaughlin(1999)]{mclaughlin99} McLaughlin, D. E. 1999, \aj,
  117, 2389

\bibitem[McLaughlin(2000)]{mclaughlin00} McLaughlin, D. E. 2000, \apj,
  539, 618

\bibitem[{McLaughlin} \& {Fall}(2008)]{mclaughlin08}
{McLaughlin}, D.~E. \& {Fall}, S.~M. 2008, \apj, 679, 1272

\bibitem[McMillan et al.(1987)]{mcmillan87} McMillan, S.~L.~W.,
  McDermott, P.~N., \& Taam, R.~E.\ 1987, \apj, 318, 261

\bibitem[McNamara, Harrison \& Baumgardt(2004)]{mcnamara04} Mcnamara,
  B.J.., Harrison, T. E., Baumgardt, H. 2004, \apj, 602, 264
  
\bibitem[{Merritt}(2009)]{merritt09} {Merritt}, D. 2009, \apj, 694,
  959

\bibitem[Metzger et al.(2009)]{metzger09}Metzger, B. D., Arcones, A.,
  Quataert, E., Martinez-Pinedo, G. 2010, \mnras, 402, 2771

\bibitem[Monaghan(1992)]{monaghan92}Monaghan, J.J. 1992, \araa, 30,
  543

\bibitem[{Moore} {et~al.}(2006)]{moore06} {Moore}, B., {Diemand}, J.,
    {Madau}, P., {Zemp}, M., \& {Stadel}, J. 2006, \mnras, 368, 563

\bibitem[Nakar et al.(2006)]{nakar06} Nakar, E. et al 2006, \apj,
  650, 281

\bibitem[Nakar(2007)]{nakarrev} Nakar, E. 2007, Physics Reports, 442,
  166

\bibitem[{Narayan} {et~al.}(1991)]{narayan91}
{Narayan}, R., {Piran}, T., \& {Shemi}, A. 1991, \apjl, 379, L17

\bibitem[{Noyola} \& {Gebhardt}(2006)]{noyola06}
{Noyola}, E. \& {Gebhardt}, K. 2006, \aj, 132, 447

\bibitem[O'Shaughnessy et al.(2005)]{osh05}O'Shaughnessy, R., Kim, C., Fragos, T., Kalogera, V., Belczynski, K., 2005, ApJ, 633, 1076

\bibitem[O'Shaughnessy et al.(2008)]{osh08}O'Shaughnessy, R., Belczynski, K., Kalogera, V., 2008, ApJ, 675, 566

\bibitem[O'Shaughnessy et al.(2010)]{osh10}O'Shaughnessy, R., Kalogera, V., Belczynski, K., 2010, ApJ, 716, 615

\bibitem[{{Paczy{\'n}ski}(1971)}]{paczynski71} {Paczy{\'n}ski}, B.
  1971, \araa, 9, 183

\bibitem[{Paczy{\'n}ski}(1986)]{paczynski86} {Paczy{\'n}ski}, B.
  1986, ApJ, 308, L43

\bibitem[{Paczy{\'n}ski \& Wiita}(1980)]{pw80} {Paczy{\'n}ski}, B., Wiita, P.J.
  1980, A\&A, 88, 23

\bibitem[Parsons, Ramirez-Ruiz \& Lee(2009)]{prrl09} Parsons, R.K.,
  Ramirez-Ruiz, E., Lee, W.H. 2009, ApJ submitted, arXiv:0904.1768

\bibitem[Pedersen et al.(2005)]{Ped05} Pedersen, K., et al.\ 2005,
  \apjl, 634, L17

\bibitem[{Phinney}(1991)]{phinney91} {Phinney}, E.~S. 1991, \apjl,
  380, L17

\bibitem[Pooley et al.(2003)]{pooley03}Pooley, D, et al., 2003, ApJ,
  591, L131

\bibitem[Press \& Teukolsky(1977)]{press77} Press, W.~H. \& Teukolsky,
  S.~A. 1977, \apj, 213, 183

\bibitem[Price \& Rosswog(2006)]{price06}Price, D. Rosswog, S. 2006,
  Science, 312, 719

\bibitem[Prochaska et al.(2006)]{prochaska06} Prochaska,
  J.~X. et~al. 2006, ApJ, 642, 989

\bibitem[{{Quinlan}(1996)}]{quinlan96}
{Quinlan}, G.~D. 1996, New Astronomy, 1, 255

\bibitem[Rasio \& Shapiro(1991)]{rass91} Rasio, F.A., Shapiro
  S.L. 1991, \apj, 377, 559

\bibitem[Rasio \& Shapiro(1992)]{rasio92} Rasio, F.A., Shapiro
S.L. 1992, \apj, 401, 226

\bibitem[Rasio \& Shapiro(1994)]{rasio94} Rasio, F.A., Shapiro
S.L. 1994, \apj, 432, 242

\bibitem[Raskin et al.(2009)]{raskin09}Raskin, C., Timmes, F.X., Scannapieco, E., Diehl, S., Fryer, C. 2009, \mnras, 399, L159

\bibitem[{Rauch} \& {Tremaine}(1996)]{rauch96} {Rauch}, K.~P. \&
  {Tremaine}, S. 1996, New Astronomy, 1, 149

\bibitem[Rees(1988)]{rees88}Rees, M.J. 1988, Nature, 333, 523

\bibitem[Rosswog et al.(2003)]{rosswog03}Rosswog, S., Ramirez-Ruiz,
E., Davis, M.B. 2003, \mnras, 345, 1077

\bibitem[Rosswog(2005)]{rosswog05} Rosswog, S. 2005, \apj, 634, 1202

\bibitem[Rosswog(2007)]{rosswog07} Rosswog, S. 2007, \mnras, 376,
  L48

\bibitem[Rosswog et al.(2009)]{rosswog09} Rosswog, S., Kasen, D., Guillochon, J., Ramirez-Ruiz, E. 2009, \apj, 705, L128

\bibitem[{Salvaterra} {et~al.}(2008)]{salvaterra08} {Salvaterra}, R.,
  {Cerutti}, A., {Chincarini}, G., {Colpi}, M., {Guidorzi}, C., \&
  {Romano}, P. 2008, \mnras, 388, L6

\bibitem[{Seth} {et~al.}(2008)]{seth08} {Seth}, A., {Ag{\"u}eros}, M.,
  {Lee}, D., \& {Basu-Zych}, A. 2008, \apj, 678, 116

\bibitem[{Shapiro}(1977)]{shapiro77}
{Shapiro}, S.~L. 1977, \apj, 217, 281

\bibitem[Shibata \& Taniguchi(2006)]{shibata06}Shibata, M.,
Taniguchi. K. 2006, Phys. Rev. D, 73, 064027

\bibitem[{Sigurdsson} \& {Phinney}(1993)]{sigurdsson93}
{Sigurdsson}, S. \& {Phinney}, E.~S. 1993, \apj, 415, 631

\bibitem[{Soderberg} {et~al.}(2006)]{soderberg06}
{Soderberg}, A.~M., {Berger}, E., {Kasliwal}, M., {Frail}, D.~A., {Price},
  P.~A., {Schmidt}, B.~P., {Kulkarni}, S.~R., {Fox}, D.~B., {Cenko}, S.~B.,
  {Gal-Yam}, A., {Nakar}, E., \& {Roth}, K.~C. 2006, \apj, 650, 261

\bibitem[Symbalisty \& Schramm(1982)]{symbalisty82}Symbalisty, E.M.D., Schramm, D.N., 1982, \apj, 1982, 22, L143

\bibitem[Usov(1992)]{usov92}Usov, V.V. 1992, Nature, 357, 472

\bibitem[van den Bosch et al.(2006)]{bosch06} van den Bosch, R., de
  Zeeuw, T., Gebhardt, K., Noyola, E., \& van de Ven, G. 2006, \apj,
  641, 852

\bibitem[Vesperini \& Heggie(1997)]{vesperini97} Vesperini, E. \&
  Heggie, D. C. 1997, \mnras, 289, 898

\bibitem[{Walcher} {et~al.}(2006)]{walcher06} {Walcher}, C.~J.,
  {B{\"o}ker}, T., {Charlot}, S., {Ho}, L.~C., {Rix}, H.-W., {Rossa},
  J., {Shields}, J.~C., \& {van der Marel}, R.~P. 2006, \apj, 649, 692

\bibitem[{Walcher} {et~al.}(2005)]{walcher05} {Walcher}, C.~J., {van
    der Marel}, R.~P., {McLaughlin}, D., {Rix}, H.-W., {B{\"o}ker},
  T., {H{\"a}ring}, N., {Ho}, L.~C., {Sarzi}, M., \& {Shields}, J.~C.
  2005, \apj, 618, 237

\bibitem[Zemp, Ramirez-Ruiz \& Diemand(2009)]{zemp09} Zemp, M., Ramirez-Ruiz, E., Diemand, J. 2009, \apj, 705, L186

\bibitem[Zheng \& Ramirez-Ruiz(2007)]{zheng07} Zheng, Z.,
  Ramirez-Ruiz, E. 2007, \apj, 665, 1220

\end{thebibliography}

\end{document}